\documentclass[11pt]{article}
\usepackage[a4paper, total={6in, 8in}]{geometry}
\usepackage{amsthm}
\usepackage{amsfonts}
\usepackage{bm}
\usepackage{graphicx}
\usepackage{natbib}
\RequirePackage[colorlinks,linkcolor=blue,citecolor=blue,urlcolor=blue]{hyperref}
\usepackage{siunitx}
\usepackage{booktabs}
\usepackage{subcaption}
\usepackage[mathscr]{euscript}
\usepackage{algorithm,algpseudocode}
\usepackage{setspace}
\usepackage{breqn}
\usepackage{mathtools}
\usepackage{multirow}
\usepackage{enumitem}
\RequirePackage{amsmath}
\usepackage{hyperref}
\usepackage{chngcntr}
\counterwithin{figure}{section}
\counterwithin{table}{section}
\newcommand{\Xset}{\bm{\mathscr{X}}} 
\newcommand{\Wset}{\bm{\mathscr{W}}} 
\newcommand{\Mset}{\bm{\mathscr{M}}} 
\newcommand{\Yset}{\bm{\mathscr{Y}}} 
\newcommand{\Natural}{\mathbb{N}} 
\newcommand{\indep}{\perp \!\!\! \perp}
\newtheorem{thm}{Theorem}[section]
\newtheorem{defn}{Definition}[section]
\newtheorem{assumption}{Assumption}
\title{A Bayesian nonparametric approach for causal inference with multiple mediators}
\vspace{0.3cm}
\author{
  Samrat Roy\\
  \small{Department of Statistics and Data Science, University of Pennsylvania}\\
   \small\textit{{email: \href{mailto:roysa@wharton.upenn.edu}{roysa@wharton.upenn.edu}}}
  \and
  Michael J. Daniels\\
  \small{Department of Statistics, University of Florida}\\
   \small\textit{{email: \href{mailto:daniels@ufl.edu}{daniels@ufl.edu}}}
  \and
  Brendan J. Kelly\\
  \small{Perelman School of Medicine, University of Pennsylvania}\\
  \small{\textit{email: \href{mailto:brendank@pennmedicine.upenn.edu}{brendank@pennmedicine.upenn.edu}}}
  \and
  Jason Roy\\
  \small{Department of Biostatistics and Epidemiology, Rutgers University}\\
  \small{\textit{email: \href{mailto:jason.roy@rutgers.edu}{jason.roy@rutgers.edu}}}
}
\begin{document}
\maketitle
\begin{abstract}
Mediation analysis with contemporaneously observed multiple mediators is an important area of causal inference. Recent approaches for multiple mediators are often based on parametric models and thus may suffer from model misspecification. Also, much of the existing literature either only allow estimation of the joint mediation effect, or estimate the joint mediation effect as the sum of individual mediator effects, which often is not a reasonable assumption. In this paper, we propose methodology which overcomes the two aforementioned drawbacks. Our method is based on a novel Bayesian nonparametric (BNP) approach, wherein the joint distribution of the observed data (outcome, mediators, treatment, and confounders) is modeled flexibly using an enriched Dirichlet process mixture with three levels: the first level characterizing the conditional distribution of the outcome given the mediators, treatment and the confounders, the second level corresponding to the conditional distribution of each of the mediators given the treatment and the confounders, and the third level corresponding to the distribution of the treatment and the confounders. We use standardization (g-computation) to compute causal mediation effects under three uncheckable assumptions that allow identification of the individual and joint mediation effects. The efficacy of our proposed method is demonstrated with simulations. We apply our proposed method to analyze data from a study of Ventilator-associated Pneumonia (VAP) co-infected patients, where the effect of the abundance of Pseudomonas on VAP infection is suspected to be mediated through antibiotics.   
\end{abstract}
\section{Introduction}\label{Intro}
Mediation analysis is an important area of causal inference. In the social, behavioral, and health sciences, including  neuroimaging ( \cite{zhao2018functional,woo2015distinct}), weight loss management (\cite{daniels2012bayesian}), air pollution (\cite{kim2019bayesian}) and  metagenomics (\cite{sohn2019compositional,wu2011linking}), researchers are often interested in estimating the part of the effect of intervention on outcome that is routed through the potential mediators. Examples include, the role of Sulfur dioxide and Nitrogen oxides emission as a mediation path between the effects of coal-fired power plants on the increase in the air pollution levels (see \cite{kim2019bayesian}). Another example (\cite{daniels2012bayesian}) relates to weight management trials, where the adherence to behavioral weight management strategies can act as a mediator between the effect of a weight management program on maintaining weight loss.   

While the early literature on mediation analysis (\cite{baron1986moderator,mackinnon1993estimating, mackinnon2008introduction}) focused on Linear Structural Equation Models (LSEM), \cite{imai2010general} formally defined the mediation effects within the counterfactual framework of causal inference and demonstrated identification under the assumption of sequential ignorability. Besides some of the other notable developments in the context of a single mediator (see \cite{frangakis2002principal}, \cite{vanderweele2009marginal}, \cite{joffe2009related}, \cite{daniels2012bayesian}), there has been recent work on multiple and high dimensional mediators. For example, \cite{sohn2019compositional} explored whether the effect of fat intake on Body Mass Index can be mediated through multiple genera of gut microbiome. 

 Recent methodological advancements have developed approaches for multiple mediators, both contemporaneously observed and causally ordered. \cite{wang2013estimation} considered estimation of mediation effects with binary outcome and contemporaneously observed multiple mediators under potential outcome framework. Their approach is based on parametric model and they define the joint mediation effect as the sum of individual mediator effects. \cite{imai2013identification}, \cite{vanderweele2014mediation} and \cite{daniel2015causal} developed methodologies which allowed for interaction effect of either contemporaneously observed, or causally related mediators. Among these, \cite{vanderweele2014mediation} confined themselves to estimation of only the joint mediation effect. On the other hand, \cite{daniel2015causal} extended the approach of \cite{imai2013identification} in the context of causally ordered mediators, providing the finest possible decomposition of the total effect into various path-specific effects. However, the proposed methodologies for estimation employ parametric approaches. In more recent times, \cite{sohn2019compositional}, in the context of contemporaneously observed compositional mediators, proposed a parametric LSEM approach, assuming no interaction effects of the mediators and thus estimated the joint mediation effect as the sum of the individual effects. To deal with high-dimensional multivariate mediators, \cite{chen2018high} developed an LSEM based approach that linearly combined the mediators into a relatively smaller number of orthogonal components, where the components are ranked based on their contribution towards the LSEM likelihood. \cite{wang2019bayesian} employed a Bayesian regularized approach to deal with both multiple exposures and mediators. However, their approach is again based on a parametric framework. Moreover, their approach allows estimation of only the joint mediation effect. Some other notable work under the parametric  paradigm includes \cite{derkach2019high}, \cite{song2020bayesian}, \cite{song2021bayesian} and \cite{zhang2021mediation}. While \cite{song2020bayesian}, \cite{song2021bayesian} and \cite{zhang2021mediation} primarily focus on mediator selection, \cite{derkach2019high} propose an approach which formalizes the mediators as latent factors. Thus, whether the mediators are observed contemporaneously or they are causally ordered, the existing literature use a parametric approach and estimate the joint mediation effect directly without any partition, or as the sum of individual mediator effects.\\
 \vspace{\baselineskip}
 
 In this paper, our objective is to develop a method that overcomes the two aforementioned drawbacks in the existing literature: (a) \textit{potential parametric misspecification} and (b) \textit{assuming that the joint mediation effect is the sum of the individual mediators effects}. We propose a novel Bayesian nonparametric (BNP) approach to causal mediation with contemporaneously observed multiple mediators, wherein, the joint distribution of the observed data (outcome, mediators, treatment, and confounders) is modeled flexibly using an extension of the enriched Dirichlet process mixture (EDPM), introduced in \cite{wade2011enriched}. In the absence of mediators, \cite{roy2018bayesian} modeled the joint distribution of outcome, treatment and confounders, using an EDPM (\cite{wade2011enriched}, \cite{wade2014improving}). We extend the EDPM to a third level: \textit{the first level} characterizing the conditional distribution of the outcome given the mediators, treatment and the confounders; \textit{the second level} corresponding to the conditional distribution of each of the mediators given the treatment and the confounders; and \textit{the third level} corresponding to the distribution of the treatment and the confounders. We introduce identifying assumptions and use standardization (g-computation; see \cite{robins1986new}, \cite{robins2009estimation} and Section \ref{comp_cau}) to compute causal mediation effects. As mentioned earlier, our method does not assume that the total mediation effect, that is, Joint Natural Indirect Effect, is the sum of the Individual Natural Indirect Effects. The proposed method is shown to have desired large sample properties.

We use our approach to analyze data from a study of critically-ill patients at risk for ventilator-associated pneumonia (VAP). VAP is a lung infection that commonly complicates the treatment of people who require a mechanical ventilator to support their breathing. The breathing tube that is placed in the upper airway to allow mechanical ventilatory support limits the clearance of lung secretions and permits bacteria that normally reside in the upper airway to increase in abundance in the lower respiratory tract. Previous literature suggests that low diversity of the respiratory bacterial microbiome or dominance of the bacterial community by a single bacterial species (a collinear measure), is associated with risk for VAP (see \cite{harrigan2021respiratory}, \cite{fernandez2020reconsidering}, \cite{ramirez2016pseudomonas}). In particular, \cite{harrigan2021respiratory} looked specifically at Pseudomonas (a bacterial genus that causes the majority of VAP cases in this population) and found that the aforementioned association varied depending on whether the subjects had recent VAP, suspected VAP, or neither. This categorization of VAP was based on the antibiotic exposure during the time interval between measuring the exposure (abundance of Pseudomonas) and measuring the outcome (VAP). Thus it would be of interest to understand the effect of individual antibiotics as mediators on the relationship between Pseudomonas dominance and VAP.

The remainder of the paper is organized as follows: in Section \ref{defn_cau}, we provide definitions of causal mediation effects and then introduce the causal identifying assumptions needed for identifiability. Section \ref{mod_bnp} introduces the BNP model for the joint distribution of the observed data. In Section \ref{Theo}, we provide some theoretical properties of our proposed method and Section \ref{comp_cau} describes the computations of the causal effects. In Section \ref{simul}, we evaluate the performance of our proposed method on simulated data and compare with some alternative approaches. We employ our approach to data from a study of Ventilator-associated Pneumonia (VAP) co-infected patients in Section \ref{application}, which is then followed by a discussion in Section \ref{dis}.           
\section{Causal Effects, Assumptions and Identifiability}
\label{defn_cau}
In this section, we define the causal mediation effects as introduced in \cite{kim2019bayesian}. Suppose we have an outcome $Y$, a binary exposure $A$, $Q$ potential mediators, $M_1, M_2, \cdots, M_Q$ that are observed contemporaneously and a vector of $p$ confounders $L$. Following the potential outcome framework introduced in \cite{imai2010identification}, we denote the potential mediators and the outcome as $M(a_1,a_2,\cdots,a_{Q})$ and $Y(a; M(a_1, a_2,\cdots, a_{Q}))$ respectively. Here $M(a_1,a_2,\cdots,a_{Q}) = \{M_1(a_1), M_2(a_2), \cdots,M_Q(a_Q)\}$ is the vector of the potential mediators, wherein $M_q(a_q)$, for $q=1,2,\cdots,Q$, corresponds to the potential value of the $q^{th}$ mediator under treatment status $a_q \in\{0,1\}$. Similarly, $Y(a; M(a_1, a_2,\cdots, a_{Q}))$ is the potential outcome under treatment status $a\in \{0,1\}$ and the mediators $M(a_1,a_2,\cdots,a_{Q})$. Thus, the potential outcome notation for $Y$ is the outcome that
corresponds to the treatment received (that is $a$) and the mediators, that are set to what they would be under different combinations of treatments $\{a_1,a_2,\cdots,a_Q\}$. Among all possible potential outcomes, only two, namely $Y(1, M(1,1,\cdots,1))$ and $Y(0, M(0,0,\cdots,0))$ are observable (and only under randomization), while all the remaining ones are (a priori) counterfactual. We define the causal effects as follows:

\vspace{0.2in}

     \noindent\textit{Total Effect} (TE): Total Effect is the entire effect of the intervention $A$ on the outcome $Y$ and is defined as the expected difference between the two observable potential outcomes, that is, $E[Y(1,M(1,1,\cdots,1))-Y(0,M(0,0,\cdots,0))]$. This total effect can be decomposed into two parts: the part that traverses directly from the intervention to the outcome, and the part that is routed through the mediators.\\
     \\
       \begin{figure}
\includegraphics[scale=1]{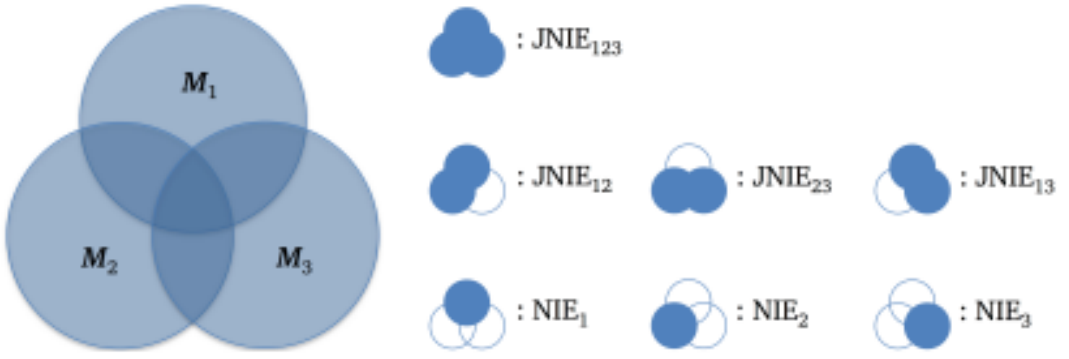}
\caption{Decomposition of Joint Natural Indirect Effect for three mediators: the decomposed effects can be attributable to any one of the mediators (individually), or, to any two of the mediators (pairwise)}
\label{fig:INIE}
\end{figure} 
\\
    \noindent\textit{Natural Direct Effect} (NDE): NDE is defined as $E[Y(1,M(0,0,\cdots,0))-Y(0,M(0,0,\cdots,0))]$. Thus this is effect of the intervention on the outcome while the mediators are set to their realizations in the absence of intervention. This effect of $A$ on $Y$ is ``direct" in the sense that it is not through the mediators.\\
    \\
    \noindent\textit{Joint Natural Indirect Effect} (JNIE): The JNIE of all $Q$ mediators is defined as $TE-NDE=E[Y(1,M(1,1,\cdots,1))-Y(1,M(0,0,\cdots,0))]$. Hence this part of the effect of $A$ on $Y$ passes indirectly through the mediators. Note that, JNIE can be decomposed into natural indirect effects that are attributable to changes in different combinations of the $Q$ mediators. In other words, JNIE comprises various indirect effects, that are mediated through different possible combinations of the $Q$ mediators. For ease of exposition, suppose $Q=3$. Then the joint effect of the three mediators $\text{JNIE}_{123}$, is depicted in Figure \ref{fig:INIE}. The JNIE can be decomposed into \textit{Individual Natural Indirect Effects} (INIE), which are attributable to any one of the three possible mediators (see $\text{JNIE}_1$, $\text{JNIE}_2$ and $\text{JNIE}_3$ in Figure \ref{fig:INIE}). For example, $\text{JNIE}_1=E[Y(1,M(1,1,1))-Y(1,M(0,1,1))]$. Similarly, one can also define \textit{Pairwise Natural Indirect Effects} (PNIE), which are passed through any two of the three mediators (see $\text{JNIE}_{12}$, $\text{JNIE}_{23}$ and $\text{JNIE}_{13}$ in Figure \ref{fig:INIE}). Thus, for example, $\text{JNIE}_{12}=E[Y(1,M(1,1,1))-Y(1,M(0,0,1))]$. For more than three mediators, the number of pairwise, three-way, etc JNIEs gets quite large; for example, for $K$ mediators, the partition of the JNIE will have $\sum {K \choose 2} + {K \choose 3} + \cdots + {K \choose K-1}$  components.  However, in applications, there are likely particular combinations that will be of primary interest.  For example, in the analysis in Section \ref{application}, there are particular pairs of mediators (antibiotics) that are frequently prescribed together.\\   
\\    
We now summarize the assumptions that are sufficient to identify the causal effects defined above.
\vspace{0.2in}
\begin{assumption}\label{A1}$\{Y(a,M(a,a,\cdots,a)), M(0,0,\cdots,0), M(1,1,\cdots,1)\} \indep A | L=\ell\}$. \end{assumption}
\vspace{\baselineskip}
\noindent The above assumption, known as \textit{Ignorable Treatment Assignment}, says that, conditional on the confounders, the treatment assignment is independent of the observable potential outcomes and the mediators. This assumption is also known as `no unmeasured confounders'.   
\vspace{0.2in}
\begin{assumption} \label{A2}
For exposure $A=1$, the conditional distributions of the observable potential outcome $Y(1,M(1,1,\cdots,1))$ given values of all potential mediators (and confounders), is the same as that of a priori counterfactual $Y(1,M(0,0,\cdots,0))$, regardless of whether the mediator values were induced by $A=1$ or $A=0$.
\end{assumption}
\vspace{\baselineskip}
\noindent Note that, the definition of JNIE involves the observable potential outcome $Y(1,M(1,1,\cdots,1))$ and a priori counterfactual $Y(1,M(0,0,\cdots,0))$. In terms of the notations, the above assumption implies that,
\begin{eqnarray*}
f_{1,M(0,0,\cdots,0)}(y\,|\,M(0,0,\cdots,0)={m}, L={\ell}) & \\
\hspace{.1in}  =  f_{1,M(1,1,\cdots,1)}(y\,|\,M(1,1,\cdots,1) ={m}, L={\ell}) . 
   \end{eqnarray*}
In other words, when the value of the mediator vector is fixed at `$m$' ( and the confounder vector is fixed at `$\ell$'), the two conditional distributions stated above are the same, irrespective of the fact that the mediators in the first case are induced in the absence of treatment, while the mediators in the second case are induced under the treatment.

The above assumption also applies to the other counterfactual outcomes, that are present in the decomposition of the JNIE. For example, for $\text{JNIE}_1$ with three mediators (see Figure \ref{fig:INIE} and relevant discussion in the definition of JNIE), the above assumption takes the following form
\begin{eqnarray*}
f_{1,M(0,1,1)}(y\,|\,M(0,1,1)={m}, L={\ell}) & \\
\hspace{.1in}  =  f_{1,M(1,1,1)}(y\,|\,M(1,1,1) ={m}, L={\ell}) . 
   \end{eqnarray*}
Similarly, for $\text{JNIE}_{12}$, the assumption translates into the following
\begin{eqnarray*}
f_{1,M(0,0,1)}(y\,|\,M(0,0,1)={m}, L={\ell}) & \\
\hspace{.1in}  =  f_{1,M(1,1,1)}(y\,|\,M(1,1,1) ={m}, L={\ell}) . 
   \end{eqnarray*}
\vspace{0.2in}   
\begin{assumption}\label{A3}
$M_{j_1}(a) \indep M_{j_2}(a^\prime) | L$ for $a \neq a^\prime$ and $j_1, j_2 =1,2,\cdots,Q$
\end{assumption}
\vspace{\baselineskip}
\noindent The above assumption says that any mediator in the absence of treatment is independent of any mediator under the treatment, conditional on the confounders $L$. This assumption is needed to identify the joint distribution of all the potential mediators. Note that, this assumption does not restrict the dependence of the mediators under the same treatment status.  
\vspace{\baselineskip}
\begin{thm}
\label{thm_ident}
Under Assumptions \ref{A1}, \ref{A2} and \ref{A3}, the NDE, JNIE ( and its decomposition) are identifiable. (see Appendix \ref{ident_proof} for the proof). 
\end{thm}
\section{BNP Model for observed data}
\label{mod_bnp}
We model the joint distribution of the outcome, mediators, treatment and confounders using an extension of the two-level enriched Dirichlet process mixture (EDPM) model (\cite{wade2011enriched}, \cite{wade2014improving}). Denoting $(A, L^T)^T$ by $X$, we propose a three-level EDP mixture model for the joint distribution of the observed data $(Y,M,X)$: 
\vspace{0.3in}
 \begin{equation}
\begin{aligned}
\label{eq:edp-med}
Y_i|M_i, X_i; \theta_i &\sim p(y|m, x;\theta_i) \\
M_{iq} | X_i; \omega_i &\sim p(m_q |x ;\omega_{i}): q=1,\ldots,Q \text{ (independent over q)}\\
X_{i,r} | \psi_i &\sim p(x_r| \psi_{i}): r=1,\ldots,p+1\text{ (independent over r)} \\
(\theta_i, \omega_i, \psi_i)|P& \sim P, \ \ \ P\sim EDP3\text{ }(\alpha_{\theta},\alpha_{\omega}, \alpha_\psi,P_0),
\end{aligned}
\vspace{\baselineskip}
\end{equation}
where, for the $i^{th}$ subject, $Y_i$, $M_i$ and $X_i$ represent the outcome, the $Q$-dimensional mediator vector and the $(p+1)$ dimensional vector containing the treatment and $p$ confounders, respectively. $M_{iq}$ is the value of the $q^{th}$ mediator for the $i^{th}$ subject and $X_{i,r}$ is the $r^{th}$ element of $X_i$. The notation, $P\sim EDP3\text{ }(\alpha_{\theta},\alpha_\omega, \alpha_\psi,P_0)$ means that
$P_\theta \sim DP(\alpha_\theta, P_{0,\theta})$, $P_{\omega |\theta}\sim DP(\alpha_\omega, P_{0,\omega|\theta})$,
and $P_{\psi|\theta, \omega}\sim DP(\alpha_\psi, P_{0,\psi|\theta, \omega})$ with base measure $P_0=P_{0,\theta}\times P_{0,\omega|\theta} \times P_{0,\psi|\theta, \omega}$, where $DP(\alpha, G)$ is a Dirichlet process with base distribution $G$ and concentration parameter $\alpha$. 

The $EDP3$, is discrete and hence, subjects can share the same values of $(\theta, \omega, \psi)$; such subjects are in the same cluster. By the construction of $P$, the aforementioned clustering is nested in three levels. The first level clusters correspond to the distinct values of the parameter $\theta$. Given a first level cluster with $\theta$ parameter as $\theta_f$, the second level clusters will correspond to those distinct values of the parameter $\omega$ for which the $\theta$ parameter is fixed at $\theta_f$. Finally, for fixed values of the parameters $(\theta, \omega)$, say $(\theta_f, \omega_s)$, the third level clusters are characterized by the distinct values of the parameter $\psi$ (see Section \ref{comp_cau} and Figure \ref{fig:cluster} for more details). The number of clusters at the three levels are controlled by the three concentration parameters, $\alpha_\theta$, $\alpha_\omega$, $\alpha_\psi$ respectively, where lower values translate into fewer clusters. The dimension of $M$ and $X$ will typically be much larger than that of $Y$. Hence, this ``enriched" three-level nested clustering, as opposed to one level clustering of a Dirichlet process, facilitates better estimation of the conditional distributions $Y\lvert (M,X)$ and $M\lvert X$ that are necessary to compute the causal effects (see Section \ref{comp_cau}).  

We assume (local / within cluster) generalized linear models for $Y_i|M_i, X_i; \theta_i$ and $M_{iq} | X_i; \omega_i$ in \eqref{eq:edp-med}. Here, given $\psi_i$, the $(p+1)$ covariates, $X_{i,1}, X_{i,2}, \cdots, X_{i,(p+1)}$ are assumed to be \textit{``locally"} (that is, intra-cluster) independent. Similarly, given $\omega_i$, and conditional on the covariates $X_i$, the $Q$ mediators $M_{i1}, M_{i2}, \cdots, M_{iQ}$ are assumed to be locally independent. This notion of local independence is similar to that in latent class models, where given latent class membership, the random variables are assumed to be independent. The assumption of local independence helps in accommodating many mediators and confounders with less computational burden, as the joint distribution is simply the product of the marginals and it does not require complex joint distributions of mediators and counfounders. However, \textit{``globally"} (that is inter-cluster) all of the variables are dependent
with potentially non-linear relationships (see various scenarios considered in the Section \ref{simul}), and this local independence can be weakened; we discuss this in Section \ref{dis}.

Note that model \eqref{eq:edp-med} can be represented using a cube-breaking formulation (\cite{wade2011enriched}), which is a generalization of the standard stick-breaking representation of DP models (\cite{sethuraman1994constructive})

$$f(y_i, m_i, x_i | P) = \sum_{j=1}^{\infty} \pi_j \text{ }p(y_i | m_i, x_i; {\theta}_j) \sum_{l=1}^\infty \pi_{l|j}  \text{ }p(m_i | x_i; \omega_{l|j}) \sum_{u=1}^\infty \pi_{u|j,l}  \text{ }p(x_i| \psi_{u|j,l}).$$\\
\vspace{\baselineskip}\\
\noindent where
    $\pi_j = \pi^\prime_j \prod_{j^\prime < j} (1-\pi^\prime_{j^\prime})$, $\pi_{l|j} = \pi^\prime_{l|j} \prod_{l^\prime < l} (1-\pi^\prime_{l^\prime|j})$ and $\pi_{u|j,l} = \pi^\prime_{u|j,l} \prod_{u^\prime < u} (1-\pi^\prime_{u^\prime|j,l})$   with 
          $\pi^\prime_j \sim \text{Beta}(1, \alpha_{\theta})$, $\pi^\prime_{l|j} \sim \text{Beta}(1, \alpha_{\omega})$ and $\pi^\prime_{u|j,l} \sim\text{Beta}(1, \alpha_{\psi})$ and $p()$ is used to denote the ``local" generalized linear models or the ``local" joint  distribution (of the covariates) mentioned above. The conditional distribution of $Y$ given $M$ and $X$ can be written as $$f(y|m,x)=\sum_{j=1}^{\infty} \Wset_j(m,x) \cdot p(y|m,x;\theta_j)$$ where, 
          $$\Wset_j(m,x)=\frac{\pi_j \sum_{l=1}^\infty \pi_{l|j}  \text{ }p(m | x; \omega_{l|j}) \sum_{u=1}^\infty \pi_{u|j,l}  \text{ }p(x| \psi_{u|j,l})}{\sum_{h=1}^{\infty}\pi_h \sum_{l=1}^\infty \pi_{l|h}  \text{ }p(m | x; \omega_{l|h}) \sum_{u=1}^\infty \pi_{u|h,l}  \text{ }p(x| \psi_{u|h,l})}.$$
\vspace{\baselineskip}\\         
Similarly, the conditional distribution of $M$ given $X$ can be written as
$$f(m|x)=\sum_{j=1}^{\infty}\sum_{l=1}^{\infty} \Wset_{j,l}(x) \cdot p(m|x;\omega_{l|j})$$ where,
$$\Wset_{j,l}(x)=\frac{\pi_j \pi_{l|j} \sum_{u=1}^\infty \pi_{u|j,l}  \text{ }p(x| \psi_{u|j,l})}{\sum_{h=1}^{\infty}\sum_{g=1}^{\infty}\pi_h \pi_{g|h} \sum_{u=1}^\infty \pi_{u|h,g}  \text{ }p(x| \psi_{u|h,g})}.$$
\vspace{\baselineskip}\\  
Note that, the the outcome model, characterized by $f(y|m,x)$, is a weighted combination of within cluster (local) regression models, with the weights as a function of $(M,X)$. Similarly, the mediator model, characterized by $f(m|x)$, is a weighted combination of within cluster regression models with the weights as a function of $X$. Thus, although the local regression models $p(y|m,x;\theta_j)$ and $p(m|x;\omega_{l|j})$, are generalized linear models, the global regression models $f(y|m,x)$ and $f(m|x)$ are computationally tractable, flexible, non-linear, non-additive models. 
\section{Theoretical Development}
\label{Theo}
In this section, we formally extend the development in \cite{wade2011enriched} to three level EDP3 introduced in the previous section. 
\vspace{0.2in}
\subsection{Three-levels Enriched Polya Urn and Enriched Dirichlet Distribution}
\label{EPU}
Consider a single urn containing $X$-balls of $k$ different colors. For each of the $i^{th}$ color of $X$-balls, there is an associated urn, denoted by $M \vert i$, that contains $M$-balls of $r$ different colors. Finally, for each of the $(i, j)^{th}$ pair corresponding to $X$-balls and $M$-balls, there is an urn, $Y \vert (j,i)$, containing $Y$-balls of $s$ different colors. More precisely, we introduce the following notation to specify the number of balls in each urn:
\begin{itemize}
\item $\alpha_i$ is the number of $X$-balls of color $i$, for $i =1,2, \cdots,k$. Thus, $\alpha(\Xset)= \sum_{i=1}^k \alpha_i$ is the total number of balls in $X$-urn
\item $\mu(j,i)$ is the number of $M$-balls of color $j$ in $M \vert i$ urn, for $j=1,2, \cdots, r$ and $i=1,2, \cdots, k$.
 Hence, $\mu(\Mset, i)= \sum_{j=1}^r \mu(j,i)$ is the total number of balls in $M \vert i$ urn. 
\item $\gamma(l,j,i)$ is the number of $Y$-balls of color $l$ in $Y \vert (j,i)$ urn, for $l=1,2, \cdots, s$, $j=1,2, \cdots, r$ and $i=1,2, \cdots, k$. Finally, $\gamma(\Yset,j,i)= \sum_{l=1}^s  \gamma(l,j,i)$ is the total number of balls in $Y \vert (j,i)$ urn.
\end{itemize}  
Using the above notation, we define the urn scheme as follows: We first draw an $X$-ball from the $X$-urn and replace it along with another ball of the same color. Then, depending on the color of the $X$-ball, draw an $M$-ball from the associated $M \vert i$ urn and replace it along with another ball of the same color. Finally, depending on the color of both $X$-ball and $M$-ball, draw an $Y$-ball from the associated $Y \vert (j,i)$ urn and replace it along with another ball of same color. We denote the result of the $n^{th}$ draw by the random vector $(X_n, M_n, Y_n)$, which is equal to $(i, j, l)$ if the $n^{th}$ $X$-ball drawn is of color $i$, $M$-ball associated with it is of color $j$ and finally the $Y$-ball associated with the previous two, is of color $l$.
\paragraph{}The above-mentioned scheme characterizes the predictive distribution as follows:
\begin{itemize}
\item $Pr(X_1=i, M_1=j, Y_1=l) = \frac{\alpha(i)}{\alpha(\Xset)} \frac{\mu(j,i)}{\mu(\Mset, i)} \frac{\gamma(l,j,i)}{\gamma(\Yset,j, i)}$ 
\item $Pr(X_{(n+1)}=i, M_{(n+1)}=j, Y_{(n+1)}=l \vert X_1=i_1, M_1=j_1, Y_1=l_1,\cdots, X_n = i_n, M_n = j_n, Y_n = l_n)=
 \frac{\alpha(i) + \sum_{h=1}^n \delta_{i_h}(i)}{\alpha(\Xset)+n} \cdot \frac{\mu(j,i) + \sum_{h=1}^n \delta_{j_h,i_h}(j,i)}{\mu(\Mset,i)+ \sum_{h=1}^n \delta_{i_h}(i)} \cdot \frac{\gamma(l,j,i) + \sum_{h=1}^n \delta_{l_h,j_h,i_h}(l,j,i)}{\gamma(\Yset,j,i)+ \sum_{h=1}^n \delta_{j_h,i_h}(j,i)}$ 
\end{itemize}
\paragraph{}
As discussed in \cite{wade2011enriched}, the above urn scheme leads to an enriched form of \textit{Dirichlet Distribution}, which is essentially a nested version of \textit{Generalized Dirichlet Distribution} of \cite{connor1969concepts}. While the basic idea remains the same as discussed in \cite{wade2011enriched}, in our case we have one more level of nesting. More specifically, in terms of notation, Equation ($4$) of \cite{wade2011enriched} now takes the following form in the context of our formulation. 
\begin{align}
&p_1,\cdots,p_k \sim Dir(\alpha(1),\cdots,\alpha(k))\\
&p_{1\vert i},\cdots,p_{r\vert i} \sim Dir(\mu(1,i),\cdots,\mu(r,i)), i=1,2,\cdots,k\\
&p_{1\vert(i,j)},\cdots,p_{s\vert(i,j)} \sim Dir(\gamma(1,j,i),\cdots,\gamma(s,j,i)), i=1,2,\cdots,k; j=1,2,\cdots,r  
\end{align}
\vspace{0.3in}
\begin{thm}\label{Theo1}
Let $\{X_n,M_n,Y_n\}_{n \in \Natural} $ be a sequence of random vectors taking values in $\{1,2,\cdots,k\} \times \{1,2,\cdots,r\} \times \{1,2,\cdots,s\}$ whose predictive distribution is characterized by a three-level Enriched Polya Urn Scheme with parameters $\alpha(\cdot)$, $\mu(\cdot,\cdot)$ and $\gamma(\cdot,\cdot,\cdot)$ as described in \ref{EPU}. Then, 
\begin{enumerate}
\item the sequence of random vectors $\{X_n,M_n,Y_n\}_{n \in \Natural} $ is exchangeable and its de Finetti measure is a three-level Enriched Dirichlet distribution with parameters $\alpha(\cdot)$, $\mu(\cdot,\cdot)$ and $\gamma(\cdot,\cdot,\cdot)$ as described in \ref{EPU}.
\item as $n\to\infty$, the sequence of predictive distributions $p_n(i,j,l)= Pr(X_{(n+1)}=i, M_{(n+1)}=j, Y_{(n+1)}=l \vert X_1=i_1, M_1=j_1, Y_1=l_1,\cdots, X_n = i_n, M_n = j_n, Y_n = l_n)$ converges a.s with respect to the exchangeable law to a random probability function, $\bm{p}$ and $\bm{p}$ is distributed according to the Enriched Dirichlet de Finetti measure.       
\end{enumerate} 
\end{thm}
\vspace{0.2in}
\subsection{Three-level Enriched Dirichlet Process and Enriched Polya Sequence}
In this section we describe the notion of the \textit{Enriched Dirichlet Process} in our setting. First we define the three-level Enriched Dirichelt Process and the Enriched Polya Sequence associated with it. Next we present some of the important properties of the three-levels Enriched Dirichlet Process, including posterior consistency. We assume that $\Xset$, $\Mset$ and $\Yset$ are complete and separable metric spaces with Borel $\sigma$-algebras $\mathbb{B}_X$, $\mathbb{B}_M$ and $\mathbb{B}_Y$. Let $\mathbb{B}$ be the $\sigma$-algebra generated by the product of the $\sigma$-algebras of $\Xset$, $\Mset$ and $\Yset$ and $\mathbb{P}(\mathbb{B})$ is the set of probability measures on the measurable product space $(\Xset \times \Mset \times \Yset, \mathbb{B})$.
\vspace{0.1in}
\begin{defn}
\begin{enumerate}
Let $\alpha$ be a finite measure on $(\Xset,\mathbb{B}_X)$. Let $\mu$ be a mapping from $(\mathbb{B}_M \times \Xset)$ to $\mathbb{R}_{+}$ such that as a function of $B \in \mathbb{B}_M$ it is a finite measure on $(\Mset,\mathbb{B}_M)$. Finally let $\gamma$ be a mapping from $(\mathbb{B}_Y \times \Mset \times \Xset)$ to $\mathbb{R}_{+}$ such that as a function of $C \in \mathbb{B}_Y$ it is a finite measure on $(\Yset,\mathbb{B}_Y)$. We then assume the following: \\
\item Law of Marginal, $Q^X$: $P_X$ is a random probability measure on $(\Xset,\mathbb{B}_X)$, where $P_X \sim DP(\alpha)$.
\item Law of conditionals, $Q_x^{M\lvert X}$ and $Q_{m,x}^{Y\lvert M,X}$: $\forall x \in \Xset, P_{M\lvert X}(\cdot\lvert x)$ is a random probability measure on $(\Mset,\mathbb{B}_M)$, where, $P_{M\lvert X}(\cdot\lvert x)\sim DP(\mu(\cdot,x))$. Similarly, $\forall x \in \Xset \text{ and } \forall m \in \Mset, P_{Y\lvert M, X}(\cdot\lvert m, x)$ is a random probability measure on $(\Yset,\mathbb{B}_Y)$, where, $P_{Y\lvert M, X}(\cdot\lvert m, x) \sim DP(\gamma(\cdot,m,x))$.
\item Joint Law of Conditionals, $Q^{M\lvert X}= \prod_{x \in \Xset}Q_x^{M\lvert X}$ and $Q^{Y\lvert M,X} = \prod_{m \in \Mset} \prod_{x \in \Xset} Q_{m,x}^{Y\lvert M,X}: P_{M\lvert X}(\cdot\lvert x), x\in \Xset$ are independent among themselves. Similarly, $P_{Y\lvert M, X}(\cdot\lvert m, x), m\in \Mset \text{ and } x\in \Xset$, are independent among themselves. 
\item Joint Law of Marginal and Conditionals, $Q= Q^X \times Q^{M\lvert X} \times Q^{Y\lvert M,X}$: $P_X$ is independent of $\{P_{M\lvert X}(\cdot\lvert x)\}_{x \in \Xset}$ and $\{P_{Y\lvert M, X}(\cdot\lvert m, x)\}_{m\in \Mset , x\in \Xset}$. Similarly, $\{P_{M\lvert X}(\cdot\lvert x)\}_{x \in \Xset}$ is independent of  $\{P_{Y\lvert M, X}(\cdot\lvert m, x)\}_{m\in \Mset , x\in \Xset}$.
\end{enumerate}
\vspace{1\baselineskip}
The joint law of marginal and conditionals, $Q$, induces the law, $\hat{Q}$, of the stochastic process $\{P(D)\}_{D\in \mathbb{B}}$ through the following reparametrization:
\begin{equation}
P(A\times B\times C)\overset{d}{=} \int_{A \times B} P_{Y\lvert M,X}(C\lvert m,x)d(P_{M\lvert X}(m\lvert x)P_X(x))
\end{equation}
for any set $A \times B \times C \in \mathbb{B}_X \times \mathbb{B}_M \times \mathbb{B}_Y$. This process is defined as a three-level Enriched Dirichlet Process with parameters $\alpha, \mu$ and $\gamma$ and is denoted by $P \sim EDP3\text{ }(\alpha,\mu, \gamma)$.  
\end{defn}
\vspace{0.1in}
\begin{defn}\label{E_pol_seq}
The sequence of random vector $\{(X_n,M_n,Y_n)\}_{n\in\mathbb{N}}$ taking values in $\Xset \times \Mset \times \Yset$ is said to be a three-level Enriched Polya Sequence with parameters $\alpha,\mu$ and $\gamma$ if:\\
\begin{enumerate}
\item For $A\in\mathbb{B}_X$ and for all $n\geq 1$,
\begin{align*}
&Pr(X_1 \in A)=\frac{\alpha(A)}{\alpha(\Xset)}\\
&Pr(X_{n+1} \in A|X_1=x_1,\cdots,X_n=x_n)= \frac{\alpha(A)+\sum_{i=1}^n\delta_{x_{i}}(A)}{\alpha(\Xset)+n}
\end{align*}
\item For $B\in\mathbb{B}_M$ and for all $n\geq 1$,
\begin{align*}
&Pr(M_1 \in B|X_1=x)=\frac{\mu(B,x)}{\mu(\Mset,x)}\\
&Pr(M_{n+1}\in B|M_1=m_1,\cdots,M_n=m_n,X_1=x_1,\cdots,X_n=x_n,X_{n+1}=x)\\[4pt]
&=\frac{\mu(B,x)+\sum_{j=1}^{n_x}\delta_{m (x,j)}(B)}{\mu(\Mset,x)+n_x}\\
\end{align*}
\item For $C\in\mathbb{B}_Y$ and for all $n\geq 1$,
\begin{align*}
&Pr(Y_1 \in C|M_1=m,X_1=x)=\frac{\gamma(C,m,x)}{\gamma(\Yset,m,x)}\\
&Pr(Y_{n+1}\in C|Y_1=y_1,\cdots,Y_n=y_n,M_1=m_1,\cdots,M_n=m_n,M_{n+1}=m,\\
&X_1=x_1,\cdots,X_n=x_n,X_{n+1}=x)\\[4pt]
&=\frac{\gamma(C,m,x)+\sum_{k=1}^{n_{mx}}\delta_{y (m x,k)}(C)}{\gamma(\Yset,m,x)+n_{mx}}\\
\end{align*}
\end{enumerate}
where $n_x= \sum_{i=1}^{n}\delta_{x_i}(x)$ and $n_{mx}= \sum_{i=1}^{n}\delta_{x_i,m_i}(x,m)$ 
\end{defn}
\paragraph{}
The three-level Enriched Polya Sequence extends the three-level Enriched Polya Urn Scheme to the case where $\Xset$, $\Mset$ and $\Yset$ are complete separable metric spaces. As discussed in \cite{wade2011enriched}, one can have similar interpretation of the above predictive distributions in Definition \ref{E_pol_seq}, in terms of draws from the urns. Also the \textit{de Finetti} measure of the above three-level Enriched Polya Urn Sequence is essentially a three-level Enriched Dirichlet Process with parameters $(\alpha, \mu, \gamma)$. Below we present some of the useful properties of the three-level Enriched Dirichlet Process.
\subsubsection{Properties}
\begin{enumerate}
\item Define $P_{0X}(\cdot)=\frac{\alpha(\cdot)}{\alpha(\Xset)}$. For every $x \in \Xset$, define $P_{0M\lvert X}(\cdot \lvert x)=\frac{\mu(\cdot,x)}{\mu(\Mset,x)}$ and finally for every $x \in \Xset$ and every $m \in \Mset$, define $P_{0Y\lvert MX}(\cdot \lvert mx)=\frac{\gamma(\cdot,m,x)}{\gamma(\Yset,,m,x)}$. Suppose $P \sim \textit{three-levels }EDP(\alpha, \mu, \gamma)$. Then from the properties of Dirichlet distribution, for every $A \in \mathbb{B}_X, B \in \mathbb{B}_M$ and $C \in \mathbb{B}_Y$ we will have the following :\\
\begin{itemize}
\item $E[P_X(A)]=P_{0X}(A)$ and $Var(P_{X}(A))=\frac{P_{0X}(A)(1-P_{0X}(A))}{\alpha(\Xset)+1}$\\
\item 
\begin{align*}
&\forall x \in \Xset, E[P_{M \lvert X}(B \lvert x)]=P_{0M \lvert X}(B \lvert x);\\ 
&Var[P_{M \lvert X}(B \lvert x)]= \frac{P_{0M \lvert X}(B \lvert x)(1-P_{0M \lvert X}(B \lvert x))}{\mu(\Mset,x)+1}
\end{align*}\\
\item 
\begin{align*}
&\forall x \in \Xset \text{ and } \forall m \in \Mset,  E[P_{Y \lvert MX}(C \lvert mx)]=P_{0Y \lvert MX}(C \lvert mx);\\ 
&Var[P_{Y \lvert MX}(C \lvert mx)]= \frac{P_{0Y \lvert MX}(C \lvert mx)(1-P_{0Y \lvert MX}(C \lvert mx))}{\gamma(\Yset,m,x)+1}
\end{align*}\\
\item $E[P(A\times B\times C)]= \int_{A \times B} P_{0Y\lvert M,X}(C\lvert m,x)d(P_{0M\lvert X}(m\lvert x)P_{0X}(x))$\\
\end{itemize}
\item If $(X_i, M_i, Y_i) \lvert P =$ \textit{P} are \textit{i.i.d}  $\sim$ \textit{P}, where $P \sim \textit{three-levels } EDP(\alpha, \mu, \gamma)$, then\\
\begin{center}
$P \lvert x_1,m_1,y_1, \cdots, x_n, m_n, y_n \sim \textit{three-levels }EDP(\alpha_n, \mu_n, \gamma_n)$
\end{center}
\begin{align*}
&\text{where} \alpha_n = \alpha + \sum_{i=1}^{n} \delta_{x_i}\\
&\\
& \forall x \in \Xset, \mu_n(\cdot,x)=\mu(\cdot,x) + \sum_{j=1}^{n_x}\delta_{m (x,j)}, \text{ where }, n_x = \sum_{i=1}^n\delta_{x_i}(x) \text{ and } \{m(x,j)\}_{j=1}^{n_x}=\{m_j : x_j=x\}\\
&\\
&\forall x \in \Xset \text{ and } \forall m \in \Mset, \gamma_n(\cdot,m,x)=\gamma(\cdot,m,x) + \sum_{k=1}^{n_{mx}}\delta_{y (mx,k)}, \text{ where }, n_{mx} = \sum_{i=1}^n\delta_{x_i,m_i}(x,m)\\ 
&\text{ and } \{y(mx,k)\}_{k=1}^{n_{mx}}=\{y_k : x_k=x, m_k=m\}\\
\end{align*}
\end{enumerate}
The following theorem provides another important property of the three-level Enriched Dirichlet Process, posterior consistency.
\begin{thm}\label{Theo_post_cons}
If $P \sim EDP3\text{ }(\alpha, \mu, \gamma)$, then for $\pi \in \mathbb{P}(\mathbb{B})$, the posterior distribution $Q_n$ of $P$ converges weakly to $\delta_{\pi}$ for $n \xrightarrow{} \infty$, a.s. $\pi^{\infty}$. 
\end{thm}
\section{Computations}
\label{comp_cau}
\begin{figure}
\includegraphics[scale=0.4]{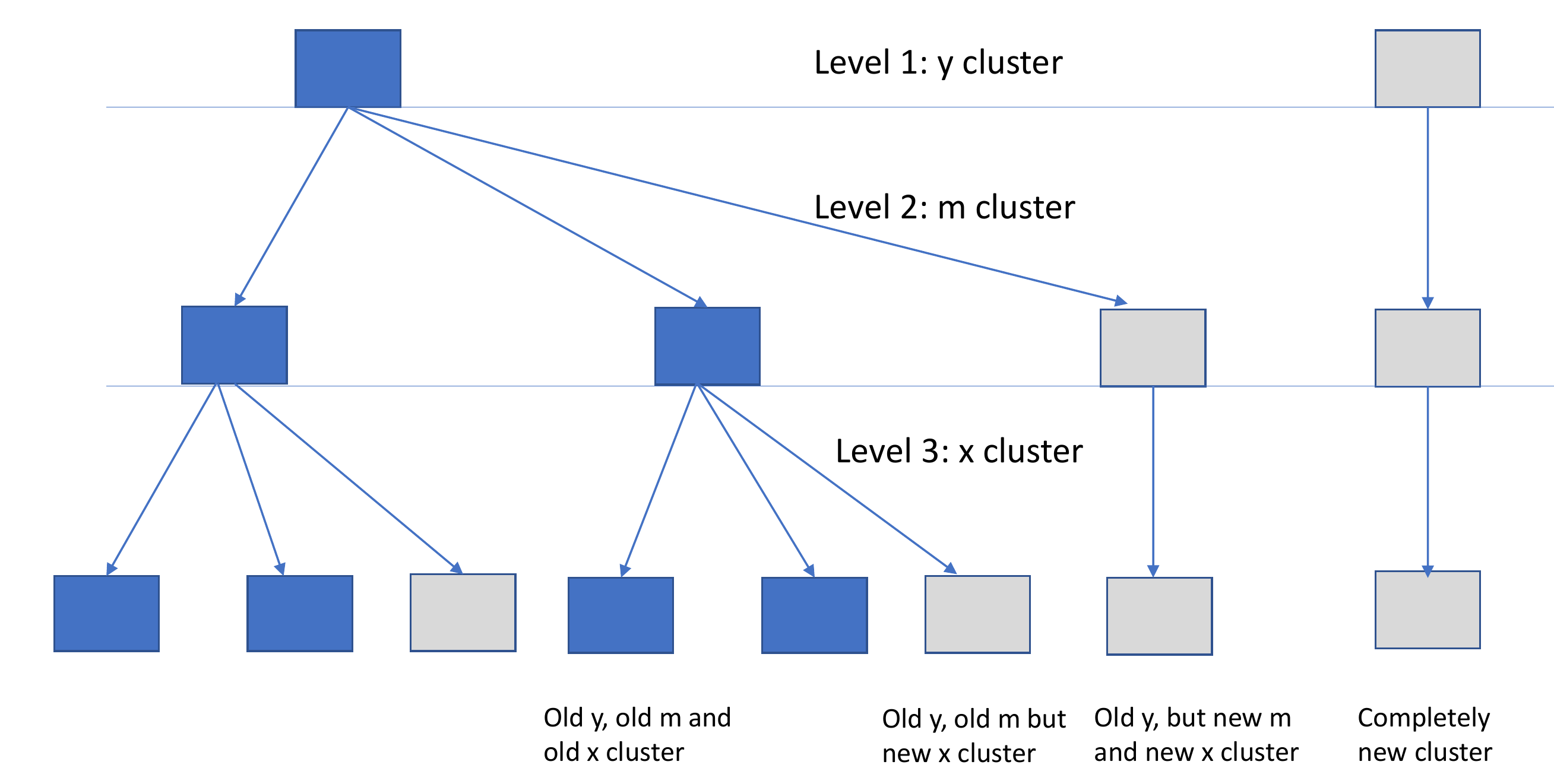}
\caption{Illustration of EDPM three level clustering}
\label{fig:cluster}
\end{figure}

We use a Gibbs Sampler to obtain draws from the posterior distributions by utilizing a further extension of Algorithm 8 in \cite{neal2000markov} from \cite{roy2018bayesian}. This algorithm accommodates nested clustering, and at each step, it alternatively samples cluster membership for each subject and then the values of the parameters, given the aforementioned cluster membership. A summary of the Gibbs Sampling is given next.
\paragraph{}Using similar notation to \cite{roy2018bayesian}, let $s_i=(s_{i,y},s_{i,m},s_{i,x})$ denote the cluster membership for the subject $i$. Here $s_{i,m}$ denotes the m-cluster within the $y$-cluster $s_{i,y}$, to which subject \textit{i} belongs. Similarly, $s_{i,x}$ characterizes the x-cluster within $(s_{i,y},s_{i,m})$ (see Figure \ref{fig:cluster}). 
We first sample $s_i$ for each subject and then given $s=\{s_i\}_{i=1}^n$, we sample the parameters $\theta$, $\omega$ and $\psi$ (see Section \ref{mod_bnp}) from their conditional distributions given the data and cluster membership. We denote by $\theta^*_j$ the $\theta$, that is associated with the $j^{th}$ currently non-empty $y$-cluster, for $j=1,2,\cdots,k$. Similarly we define $\omega^*_{l \vert j}$ and $\psi^*_{u \vert j,l}$ for $l=1,2,\cdots k_j$ and $u=1,2,\cdots k_{jl}$, where $k_j$ is the number of currently non-empty $m$-clusters within the $j^{th}$ $y$-cluster and $k_{jl}$ is the number of currently non-empty $x$-clusters within the $j^{th}$ $y$-cluster and $l^{th}$ $m$-cluster. Given these draws of the cluster membership and the parameters, we perform the post-processing computations of the causal effects defined in Section \ref{defn_cau}. For the purpose of demonstration, it is sufficient to describe the computation of a generic expected potential outcome $E[Y(a,M(a_1,a_2,\cdots,a_Q))]$, since all the causal effects defined in Section \ref{defn_cau}, involve $E[Y(a,M(a_1,a_2,\cdots,a_Q))]$ for different combinations of $\{a,a_1,a_2,\cdots,a_Q\} \in \{0,1\}$.
\paragraph{}
Given the cluster membership and cluster-specific parameters, we conduct the following post-processing steps to compute $E[Y(a,M(a_1,a_2,\cdots,a_Q))]$ for that posterior sample of cluster membership and parameters.  
\begin{itemize}
    \item[(a)] Draw the covariates $l$ (see Step a in Appendix \ref{detailed_comp})\\
    \item[(b)]Given the covariates from step (a), draw the mediators $m$ in such a way that the $q^{th}$ mediator is induced under the treatment status $a_q$, for some fixed set $\{a_1,a_2, \cdots,a_Q\}\in\{0,1\}$ (see Step b in Appendix \ref{detailed_comp})\\
    \item[(c)] Given the values from (a) and (b), compute $E(Y \vert A=a, L=l, M=m, \theta^*, \omega^*, \psi^*,s)$ (see Step c in Appendix \ref{detailed_comp}). $\theta^*$, $\omega^*$ and $\psi^*$ are used to denote $\{\theta^*_j\}_{j}$, $\{\omega^*_{l\lvert j}\}_{j,l}$ and $\{\psi^*_{u\lvert j,l}\}_{j,l,u}$ respectively from the particular posterior sample and $s=\{s_i\}_{i=1}^n$ denotes the corresponding cluster memberships.\\
    \item[(d)]Repeat steps (a)-(c) T times and use Monte Carlo Integration to compute $E(Y(a,M(a_1,a_2,\cdots,a_Q)))$ (see Step d in Appendix \ref{detailed_comp}).\\
\end{itemize}

\paragraph{} Using the above steps, one can compute any of the expected potential outcomes required to construct the causal effects. It is worth mentioning that the above-mentioned computation is a post-processing step, that can be done outside the Gibbs Sampler. Hence this step can easily be parallelized.  
\section{Simulation Studies}
\label{simul}
In this section, we evaluate the performance of our proposed methodology on synthetic data under different scenarios. The true data generating mechanism and the numerical results for each of the scenarios are discussed next. For all scenarios, we report point estimates (posterior means), $95\%$ CI widths and empirical coverage probabilities corresponding to the estimation of NDE, JNIE, TE and compare the results with those obtained using an LSEM approach with bootstrap. We also present results on Individual Natural Indirect Effects (INIE), and for Scenario 3, we additionally report the results on Pairwise Natural Indirect Effects for some pairs of mediators. Also, note that the data generating mechanism in Scenario 6 is in line with the real data considered in Section \ref{application}. Some of the tables related to this section are deferred to Appendix \ref{A3}.                
\vspace{0.4cm}
\begin{table*}
\centering
\begin{tabular}{c c c c c}
    \multicolumn{2}{c}{}& \multicolumn{1}{c}{\textit{Estimate}} & \multicolumn{1}{c}{\textit{CI Width}} & \multicolumn{1}{c}{\textit{Coverage}}\\  
    \hline
      \multirow{2}{*}{\textit{True NDE}=$1.04$} & \small{BNP ($n=1000$)} &1.04&0.59&0.96\\
      &\small{BNP ($n=2000$)} &1.04&0.47&0.98\\
      &\small{LSEM ($n=1000$)} &1.04&0.38&0.93\\
      &\small{LSEM ($n=2000$)} &1.04&0.35&0.93\\
      \hline
      \multirow{2}{*}{\textit{True JNIE}=$0.71$} & \small{BNP ($n=1000$)} &0.70&0.83&1.0\\
      &\small{BNP ($n=2000$)} &0.70&0.68&1.0\\
      &\small{LSEM ($n=1000$)} &0.70&0.61&0.94\\
      &\small{LSEM ($n=2000$)} &0.70&0.55&0.95\\
       \hline
       \multirow{2}{*}{\textit{True TE}=$1.75$} & \small{BNP ($n=1000$)} &1.74&0.85&1.0\\
       &\small{BNP ($n=2000$)} &1.74&0.69&1.0\\
      &\small{LSEM ($n=1000$)} &1.74&0.70&0.93\\
      &\small{LSEM ($n=2000$)} &1.74&0.66&0.94\\
       \hline
    \end{tabular}
    \caption{Scenario 1 results for NDE, JNIE, TE}
    \vspace{0.8cm}
    \label{Sim_sc1}
\end{table*}

\noindent\textbf{Scenario 1}: \textit{Continuous outcome and mediators, simple functional forms}\\
\noindent We generate a covariate matrix $L \in \mathbb{R}^{n \times (p_1+p_2)}$, wherein the first $p_1$ columns contain the discrete covariates and the remaining $p_2$ columns are for continuous covariates. Discrete covariates are generated independently from Bernoulli $(p=0.5)$. The continuous covariates are generated independently of the discrete covariates and are drawn from $N_{p_2}(0,\Sigma_c)$, where $\Sigma_c \in \mathbb{R}^{p_2 \times p_2}$ is the covariance matrix of the $p_2$-variate Normal distribution, for which all the off-diagonal elements are $0.3$. The vector of binary treatment $A \in \mathbb{R}^{n}$ is generated from Bernoulli (prob=0.4). The mediator matrix is denoted by $M \in \mathbb{R}^{n \times Q}$, for which the columns correspond to the $Q$ mediators and they are generated independently as follows:
    \vspace{0.3cm}
    \begin{equation}
    \label{nkernel_1}
        M_q \lvert A,L \sim \gamma_m\text{ } N(\mu_{m1},\mathbb{I}_n) + (1-\gamma_m) N(\mu_{m2},\mathbb{I}_n), \text{ for } q=1,2,\cdots,Q
        \vspace{0.3cm}
    \end{equation}
    where, $\gamma_m$ is a vector of length $n$, which is generated as $n$ i.i.d. samples from Bernoulli($\delta_m$), where $\delta_m$ is taken as 0.4. The means of the Normal distributions are taken as $\mu_{m1}=-4+2A-0.5L_{p_1+2}-L_{p_1+3}+0.5L_{p_1+4}$ and $\mu_{m2}=-4+0.4A+0.5L_{p_1+2}-0.8L_{p_1+3}$.
    
    Finally, the outcome vector $Y \in \mathbb{R}^n$ is generated as follows:
    \vspace{0.3cm}
    \begin{equation}
    \label{nkernel_2}
        Y \lvert M,A,L \sim \eta_y\text{ } N(\mu_{y1},\mathbb{I}_n) + (1-\eta_y) N(\mu_{y2},\mathbb{I}_n)
        \vspace{0.3cm}
    \end{equation}
    where, $\eta_y$ is generated as $n$ i.i.d. samples from Bernoulli($\zeta_y$), with $\zeta_y\text{ }=0.4$. $\mu_{y1}=-4+2A-0.5L_{p_1+2}+0.5M_{Q}$ and $\mu_{y2}=-2+0.4A+0.5L_{p_1+2}+0.8M_{Q}$. We set $p_1$ and $p_2$ to $4$, $Q=10$ and obtain the results for $n=1000$ and $2000$. Note that, in this scenario, both the mediator model and the outcome model involve simple linear functional forms and they do not include any nonlinear or interaction terms. Thus, in terms of structural complexity, this is the simplest scenario that we consider. Due to this simple structure, the LSEM approach should perform well, especially, since we are using bootstrap to quantify the uncertainty. Table \ref{Sim_sc1} summarizes the results for NDE, JNIE and TE using our BNP approach and the LSEM approach. As expected, both the LSEM approach and our BNP approach perform well in terms of bias and coverage, though BNP shows over-coverage in some of the cases, and LSEM has smaller CI widths. Table \ref{scn_1_INIE} shows the results for INIE of each of the mediators for $n=1000$. Note that the outcome model involves only the $Q^{th}$ mediator $M_Q$. Thus, out of all 10 mediators, only the last one (that is the $Q^{th}$ one, $Q=10$) acts as a true mediator here. As it can be seen from Table \ref{scn_1_INIE}, the INIE estimation for the true mediator is quite good in terms of both bias and coverage.    
    \vspace{0.3cm}

    \noindent \textbf{Scenario 2}: \textit{Continuous outcome and mediators, complex functional forms involving nonlinear and interaction terms}\\
    \noindent In Scenario 2, the steps of generating the data remain the same as in Scenario 1. However, here we introduce more complex functional forms involving nonlinear and interaction terms. To that end, while generating A, the Bernoulli probability now has the form $\text{logit}^{-1}(0.3\sum_{j=1}^4L_{p_1+j})$. The means of the Normal distributions are considered as $\mu_{m1}=-4+2A-0.5L_{p_1+2}-L_{p_1+3}+0.5L_{p_1+4}$, $\mu_{m2}=4+0.4A+0.5L_{p_1+2}^2-0.8L_{p_1+3}(L_{p_1+3}>0)$,
    $\mu_{y1}=-4+2A-0.5L_{p_1+2}* M_Q-L_{p_1+3}*M_Q+0.5L_{p_1+4}*M_Q$ and $\mu_{y2}=4+0.4A+0.5L_{p_1+2}^2-0.8L_{p_1+3}(L_{p_1+3}>0)$. Finally, while obtaining the mixing probabilities $\gamma_m$ and $\eta_y$, the corresponding Bernoulli parameters $\delta_m$ and $\zeta_y$ are specified as follows:
    \vspace{0.3cm}
    $$
        \delta_m= \frac{exp\{-2(L_{p_1}+1)^2\}}{exp\{-2(L_{p_1}+1)^2\}+exp\{-2(L_{p_1}-2)^2\}}
    $$
    and 
    $$
        \zeta_y= \frac{exp\{-2(M_Q+1)^2\}}{exp\{-2(M_Q+1)^2\}+exp\{-2(M_Q-2)^2\}}
        \vspace{0.4cm}
    $$

    The results are summarized in Table \ref{Sim_scn2} (NDE, JNIE and TE) and Table \ref{scn_2_INIE} (INIE). As shown in Table \ref{Sim_scn2}, the NDE and JNIE results based on LSEM approach suffer from parametric misspecification and the BNP approach significantly outperforms LSEM, though the credible intervals for BNP are a bit conservative. The INIE results, summarized in Table \ref{scn_2_INIE}, are very good for the true mediator (the $Q^{th}$ mediator, $Q=10$; see Scenario 1 for more details) and moreover, they show improvement for increased sample size (Table \ref{scn_22_INIE} in Appendix \ref{add_tables}).

    \begin{table*}
\centering
\begin{tabular}{c c c c c}
    \multicolumn{2}{c}{}& \multicolumn{1}{c}{\textit{Estimate}} & \multicolumn{1}{c}{\textit{CI Width}} & \multicolumn{1}{c}{\textit{Coverage}}\\  
    \hline
      \multirow{2}{*}{\textit{True NDE}= $1.49$} & 
      \small{BNP ($n=1000$)} &1.43&1.07&0.87\\
      &\small{BNP ($n=2000$)} &1.50&0.92&0.93\\
      &\small{LSEM ($n=1000$)} &2.40&0.64&0.00\\
      &\small{LSEM ($n=2000$)} &2.39&0.44&0.00\\
      \hline
      \multirow{2}{*}{\textit{True JNIE}= $4.61$} & 
      \small{BNP ($n=1000$)} &5.40&2.20&0.88\\
      &\small{BNP ($n=2000$)} &5.21&1.97&0.92\\
      &\small{LSEM ($n=1000$)} &3.81&1.71&0.62\\
      &\small{LSEM ($n=2000$)} &3.82&1.21&0.63\\
       \hline
       \multirow{2}{*}{\textit{True TE}= $6.13$} & 
       \small{BNP ($n=1000$)} &6.31&2.17&0.87\\
       &\small{BNP ($n=2000$)} &6.24&1.85&0.93\\
      &\small{LSEM ($n=1000$)} &6.98&1.84&0.92\\
      &\small{LSEM ($n=2000$)} &6.75&1.43&0.93\\
       \hline
    \end{tabular}
    \caption{Scenario 3 results for NDE, JNIE, TE}
    \vspace{0.4cm}
    \label{Sim_scn3}
\end{table*}
\vspace{0.4cm}
\vspace{0.4cm}
\begin{table}
\centering
\begin{tabular}{|c | c| c| c| c|}
\hline
\textit{Mediator No.} & \textit{True INIE} & \textit{Estimate} & \textit{CI Width} & \textit{Coverage}\\
\hline
1&0 &0.03 &1.11 &0.98 \\
2&0& 0.02&1.21 &0.99 \\
3&0&0.02 & 1.13& 0.98\\
4&0&0.03 &1.18 &0.99 \\
5&0&0.024 &1.12 &0.98 \\
6&0&0.01& 1.09&0.99 \\
7&0& 0.02& 1.21&0.98 \\
8& -2.47&-2.13 &1.28 &0.96 \\
9&2.65&2.33&1.13&0.97 \\
10&5.68&5.11 &1.14 &0.98 \\
\hline
\end{tabular}
\caption{Scenario 3 INIE results for $n=1000$} \label{scn_3_INIE}
\end{table}
\begin{table}
\centering
\begin{tabular}{|c | c| c| c| c|}
\hline
\textit{Mediator pairs} & \textit{True PNIE} & \textit{Estimate} & \textit{CI Width} & \textit{Coverage}\\
\hline
8 \textit{and} 9&1.56&1.32& 1.12&0.98 \\
8 \textit{and} 10&3.23&2.99&1.22&0.99 \\
9 \textit{and} 10&5.71&5.46&1.16&0.97\\
\hline
\end{tabular}
\caption{Scenario 3 Pairwise NIE (PNIE) results for $n=1000$} \label{scn_3_PNIE}
\end{table}
    \noindent \textbf{Scenario 3}: \textit{Continuous outcome and mediators, mediators are correlated and outcome model involves interaction terms among the mediators}\\
    \noindent In Scenario 3, the data generating steps are mostly in line with that of Scenario 2. However, here we induce correlation among the $Q$ mediators and also include interaction among the mediators in the outcome model. As defined in Scenario 1, each row of the mediator matrix $M \in \mathbb{R}^{n \times Q}$ corresponds to a particular individual or observation. Thus to induce correlation among the mediators, we generate the rows of $M$, denoted by $\{M^i\}_{i=1}^n$ independently as follows:
    \vspace{0.4cm}
    $$
        M^i \lvert A,L \sim \gamma_m^i\text{ } N(\mu_{1}^i,\Sigma_M) + (1-\gamma_m^i) N(\mu_{2}^i,\Sigma_M), \text{ for } i=1,2,\cdots,n
        \vspace{0.4cm}
    $$
    where $\gamma_m^i$ is the $i^{th}$ element of $\gamma_m$ and $\mu_{1}^i=\{\mu_{m1}^i, \mu_{m1}^i, \cdots, \mu_{m1}^i\}^T \in \mathbb{R}^Q$ and $\mu_{m1}^i$ is the $i^{th}$ element of $\mu_{m1}$. $\mu_{2}^i$ is defined as $\mu_1^{i}$ and $\Sigma_M\in\mathbb{R}^{Q \times Q}$ represents the covariance structure among the mediators, whose diagonal elements are 1 and the off-diagonal elements are 0.45. Finally, to introduce the interaction terms among the mediators, the means of the outcome model are taken as $\mu_{y1}=-4+2A-0.5L_{p_1+2}* M_Q-M_{Q-1}*M_Q+0.5M_{Q-2}*M_{Q-1}$ and $\mu_{y2}=4+0.4A+0.3M_{Q-2}*M_{Q}-0.8L_{p_1+3}(L_{p_1+3}>0)$. This scenario is more difficult than the previous one for the following reasons. First, the outcome model considered here includes the interaction among the mediators. Thus, as opposed to the previous two scenarios, here we have three true mediators, namely, mediators 8, 9 and 10 (that is, $M_{Q-2}$, $M_{Q-1}$ and $M_{Q}$ for $Q=10$). In addition to that, this scenario allows the mediators to be correlated. Table \ref{Sim_scn3} summarizes the NDE, JNIE and TE results under both BNP and LSEM approach. As expected, the LSEM results again suffer from parametric misspecification and BNP significantly outperforms LSEM. For example, the estimated NDE for BNP with $n=1000$ is $1.43$ which is quite close to the true NDE $1.49$, as compared to the LSEM estimate $2.40$. For JNIE, though the performances are similar in terms of bias (the estimates are $5.40$ and $3.81$ for BNP and LSEM respectively, while the true value is 4.61), the coverage is significantly worse for LSEM. Note that, for this scenario, the BNP coverage values are slightly lower than the earlier scenarios. This is potentially due to the fact that the mediators are correlated here, while BNP assumes local independence (intra-cluster independence, see Section \ref{mod_bnp}) of the mediators. However, even under this difficult scenario, BNP performs considerably well and the coverage approaches the desirable level as we increase the sample size. The INIE results, summarized in Table \ref{scn_3_INIE}, are also quite good for the true mediators (mediators $8$, $9$ and $10$) and moreover, as shown in Table \ref{scn_33_INIE} of Appendix \ref{add_tables}, the bias for INIE decreases as we increase the sample size. Since this scenario has more than one true mediator, we report the results for pairwise indirect effects (PNIE). As summarized in Table \ref{scn_3_PNIE}, the results for PNIE are good in terms of both bias and coverage.           
    \vspace{0.4cm}

    \noindent \textbf{Scenario 4}: \textit{Continuous outcome and mediators, complex functional forms and skewed errors}\\
    \noindent Scenario 4 is similar to Scenario 2, except for the distributional assumption of the errors. To allow for skewed error, we replace the normal kernels in Equations (\ref{nkernel_1}) and (\ref{nkernel_2}) with the Skew Normal distribution (\cite{evans2001statistical}) The location parameters are $\mu_{m1}$, $\mu_{m2}$, $\mu_{y1}$ and $\mu_{y2}$ and the scale parameters are taken as 1. The shape parameters, that characterize the skewness of the distribution, are taken as $\alpha=4$. The results are summarized in Tables \ref{Sim_scn4} and \ref{scn_4_INIE}. Similar to Scenario 2, BNP significantly outperforms LSEM.
    
\vspace{0.4cm}
 
    \noindent \textbf{Scenario 5}: \textit{Continuous outcome and mediators, complex functional forms and correlated binary covariates}\\
    \noindent Scenario 5 is also similar to Scenario 2, except for the fact that here we induce correlation among the binary covariates $L_1$, $L_2$, $\cdots, L_{p_1}$ using a Gaussian copula with a correlation matrix whose off-diagonal elements are $0.6$. Table \ref{Sim_scn5} and Table \ref{scn_5_INIE} summarize the results. The results show that BNP significantly outperforms LSEM and also the INIE results for the true mediator (the $Q^{th}$ mediator, $Q=10$) are quite good in terms of bias and coverage.
\vspace{0.6cm}\\
\noindent \textbf{Scenario 6}: \textit{Binary outcome, binary mediators, and binary confounders, complex functional forms}
   \begin{table*}
\centering
\begin{tabular}{c c c c c}
    \multicolumn{2}{c}{}& \multicolumn{1}{c}{\textit{Estimate}} & \multicolumn{1}{c}{\textit{CI Width}} & \multicolumn{1}{c}{\textit{Coverage}}\\  
    \hline
      \multirow{2}{*}{\textit{True NDE}=0.29} & \small{BNP ($n=100$)} &0.25&0.35&0.92\\
      &\small{BNP ($n=300$)} &0.28&0.25&0.96\\
      \hline
      \multirow{2}{*}{\textit{True JNIE}= -0.02} & \small{BNP ($n=100$)} &-0.02&0.39&0.95\\
      &\small{BNP ($n=300$)} &-0.02&0.23&0.97\\
       \hline
       \multirow{2}{*}{\textit{True TE}=0.27} & \small{BNP ($n=100$)} &0.23&0.51&0.97\\
       &\small{BNP ($n=300$)} &0.26&0.34&0.99\\
       \hline
    \end{tabular}
    \caption{Scenario 6 results for NDE, JNIE, TE}
    \vspace{0.8cm}
    \label{Sim_scn6}
\end{table*}
\vspace{0.4cm}

\noindent We generate the data in accordance with the real data analysis in Section \ref{application}. First we generate a binary covariate matrix $L \in \mathbb{R}^{n \times p_1}$, wherein the $p_1$ columns are generated independently of each other, where each of them are $n$ i.i.d. samples from Bernoulli $(p=0.5)$. The vector of binary treatment $A \in \mathbb{R}^{n}$ is obtained as $n$ i.i.d. samples from Bernoulli (prob=0.4). The binary mediator matrix is denoted by $M \in \mathbb{R}^{n \times Q}$, for which the columns correspond to the $Q$ mediators and they are generated independently as follows:
    \vspace{0.3cm}
    $$
        M_q \lvert A,L \sim \gamma_m\text{ } Bernoulli\text{ }(\Phi(\mu_{m1})) + (1-\gamma_m) \text{ } Bernoulli\text{ }(\Phi(\mu_{m2})), \text{ for } q=1,2,\cdots,Q
        \vspace{0.3cm}
    $$
    where, $\gamma_m$ is a vector of length $n$, which is generated as $n$ i.i.d. samples from Bernoulli($\delta_m$), where $\delta_m$ is taken as 0.4. The success probabilities for the Bernoulli distributions are $\Phi(\mu_{m1})$ and $\Phi(\mu_{m2})$, with $\mu_{m1}=-1+3A+0.5L_{p_1-2}+L_{p_1-1}+0.5L_{p_1}$ and $\mu_{m2}=-2+A+0.5L_{p_1-2}^2-0.8L_{p_1-1}(L_{p_1}>0)$.
    
    Finally, the outcome vector $Y \in \mathbb{R}^n$ is generated as follows:
    \vspace{0.3cm}
    $$
        Y \lvert M,A,L \sim \eta_y\text{ } Bernoulli\text{ }(\Phi(\mu_{y1})) + (1-\eta_y)\text{ }Bernoulli\text{ }(\Phi(\mu_{y2}))
        \vspace{0.3cm}
    $$
    where $\eta_y$ is generated as $n$ i.i.d. samples from Bernoulli($\zeta_y$), with $\zeta_y\text{ }=0.4$ and $\mu_{y1}=-4+5A-0.5L_{p_1-2}*M_{Q}-L_{p_1-1}*M_{Q}+0.5L_{p_1}*M_{Q}$ and $\mu_{y2}= 4+6A+0.5L_{p_1-2}^2-0.8L_{p_1-1}(L_{p_1}>0)$. We take $p_1$ as $4$ and $Q=10$ as before. To be consistent with the sample size in real data analysis in this scenario, we obtain the results for $n=100$ and $n=300$. Table \ref{Sim_scn6} summarizes the results for NDE, JNIE and TE, while Table \ref{scn_6_INIE} shows the results for INIE of each of the mediators. As seen from Tables \ref{Sim_scn6} and \ref{scn_6_INIE}, our approach performs well in terms of both bias and coverage.
\section{Application}
\label{application}
In this section we apply our proposed method to data on \textit{Ventilator-associated Pneumonia} (VAP) co-infected patients. VAP is a lung infection  that can develop in a person who is on a ventilator. An infection may occur if germs enter through the tube and get into the patient’s lungs. As discussed in Section \ref{Intro}, existing literature have showed that low diversity of the respiratory bacterial microbiome, or dominance of the bacterial community by a single bacterial species (a collinear measure), is associated with risk for VAP (see \cite{harrigan2021respiratory}, \cite{fernandez2020reconsidering}, \cite{ramirez2016pseudomonas}), Their findings also suggested that it would be of interest to understand the effect of individual antibiotics as mediators of the relationship between Pseudomonas dominance and VAP (see Section \ref{Intro} for more details). 

Our data consists of 83 hospitalized subjects on mechanical ventilation, and at risk for VAP. The exposure is a binary variable that characterizes abundance of Pseudomonas at presentation. The outcome takes the value 1 if the subject gets infected by Pseudomonas VAP within 30 days. The study considers 15 different antibiotics: Vancomycin IV, Metronidazol ,Cefazolin, Daptomycin, Linezolid, Meropenem, Pip Tazo, Cefepime, Levofloxacin, Colistimethate, Clindamycin, Ceftriaxone, Azithromycin, Ampsul and Amikacin. The mediators are defined as whether that particular antibiotic was on or off during the $30$ day window between exposure and outcome. Four medical comorbidities, COPD, Asthma, ILD (Interstitial Lung Disease) and Lymphoma/leukemia were considered as confounders.
\begin{table}
  \centering
  \begin{tabular}{c|c|c|c}
    \toprule
    Causal Effect & Estimate & Lower CI& Upper CI\\
    \midrule
    NDE&0.26&0.08&0.45\\
JNIE&-0.05&-0.31&0.21\\
TE&0.2&-0.1&0.47\\
\bottomrule
\end{tabular}
\caption{NDE, JNIE and TE: Posterior means and $95\%$ Credible Intervals}
\label {data_ana}
\end{table}

Table \ref{data_ana} summarizes the results for NDE, JNIE and TE, and Table \ref{data_ana_tab_2} summarizes the results for individual mediator effects. As seen from Table \ref{data_ana}, the joint indirect effect of the antibiotics is very small. Also, as Table \ref{data_ana_tab_2} shows, none of the individual effects are important. We also estimate the pairwise natural indirect effects for the pairs of antibiotics often prescribed together: (Vancomycin IV, Pip Tazo), (Vancomycin IV, Cefepime), (Vancomycin IV, Levofloxacin), (Daptomycin, Pip Tazo), (Daptomycin, Cefepime), and (Daptomycin, Levofloxacin), and the results are summarized in Table \ref{data_ana_tab_3}. These pairs also do not show any mediation effect. Note that, the CI widths for INIEs (and PNIEs) in Table \ref{data_ana_tab_2} (and Table \ref{data_ana_tab_3}) are very similar. To confirm that the data was informing on these, we created similar CI's by sampling the regression parameters from their prior distributions; the resulting CI's were much wider.               
\begin{table}
  \centering
  \begin{tabular}{c|c|c|c}
    \toprule
    Mediator & INIE Estimate & Lower CI& Upper CI\\
    \midrule
    Vancomycin IV, Pip Tazo&0.02&-0.20&0.25\\
    Vancomycin IV, Cefepime&0.02&-0.19&0.24\\
    Vancomycin IV, Levofloxacin&0.03&-0.21&0.23\\
    Daptomycin, Pip Tazo&0.01&-0.19&0.22\\
    Daptomycin, Cefepime&0.01&-0.20&0.23\\
    Daptomycin, Levofloxacin&0.01&-0.20&0.22\\
\bottomrule
\end{tabular}
\caption{Pairwise-NIE for the selected pairs of mediators: Posterior means and $95\%$ Credible Intervals}
\label{data_ana_tab_3}
\end{table}

\section{Discussion}
Mediation analysis with contemporaneously observed mediators is an important area of causal inference with wide applicability. Existing methodologies dealing with multiple mediators are typically based on parametric models and thus may suffer from parametric misspecification. In addition, the existing literature estimates the joint mediation effect as the sum of individual mediators effect, which, in most settings, is not a reasonable assumption. In this paper, we proposed a methodology which overcomes the two aforementioned drawbacks. Our method is based on a novel Bayesian nonparametric (BNP) approach, which modeled the joint distribution of the observed data (outome, mediators, treatment, and confounders) flexibly, using an enriched Dirichlet process mixture with three levels. The first level specified the conditional distribution of the outcome given the mediators, treatment and the confounders, the second level characterized the conditional distribution of each of the mediators given the treatment and the confounders, and the third level characterized the distribution of the treatment and the confounders. Causal effects were identified under some suitable causal assumptions and the proposed method was shown to have desired large sample properties. The efficacy of our proposed method was demonstrated with simulations. We applied our proposed method to analyze data from a study of Ventilator-associated Pneumonia (VAP) co-infected patients, where the effect of the abundance of Pseudomonas on VAP infection was suspected to be mediated through the antibiotics. However, both the joint and individual mediation effects of the antibiotics were not significant. Some future directions of this work are as follows. First, our approach assumes ``local" independence among the covariates and among the mediators conditional on the covariates (see Section \ref{mod_bnp}). As an extension, it might be of interest to induce ``local" dependence, in particular among the mediators. For binary mediators, the dependence can be induced through the Dirichlet process mixture of the product multinomials (see \cite{dunson2009nonparametric}), as the mediators can be considered as unordered categorical variables. Note that this specification makes it difficult to allow the mediators to depend on the covariates locally. Another potential future work could extend our approach to accommodate mediator variable selection. In particular, one might be interested in checking whether a particular mediator (say, the $j^{th}$ mediator) is independent of the outcome, conditional on the exposure, confounders and the other mediators (i.e., $M_j \perp Y \lvert A, L, M_{-j}$) and also whether the mediator is independent of the exposure, conditional on the confounders and the other mediators (i.e., $M_j \perp A \lvert L, M_{-j}$). As opposed to the parametric approach, our nonparametric specification does not have any parameters corresponding to the aforementioned conditional independencies and thus it requires a different way to perform mediator selection (Dhara, Daniels and Roy 2022 (working paper)). Finally, it would be of interest to perform sensitivity analysis to the causal assumptions, in particular Assumption \ref{A3}. However, performing sensitivity analysis on Assumption \ref{A3} is tricky since there is not an implicit covariance matrix to fully specify the joint distribution of $M(0)$ and $M(1)$.\\
\label{dis}
\vspace{\baselineskip}
\\
\textbf{\scalebox{1.3}{Appendix}}
\appendix
\section{Computational Details for Post-processing}
\label{detailed_comp}
In this section, we provide a detailed description of the post-processing computation steps for $E(Y(a,M(a_1,a_2,\cdots,a_Q)))$, summarized in Section \ref{comp_cau}. For each posterior sample, 
\vspace{0.1in}\\
\begin{itemize}
    \item \textit{\textbf{Step a}}\\ 
     \textit{Substep a.1}: Draw a first level cluster from a Multinomial $\{1,2,\cdots,k+1\}$ with probabilities $(\frac{n_1}{\alpha_{\theta}+n},\frac{n_2}{\alpha_{\theta}+n},\cdots,\frac{n_k}{\alpha_{\theta}+n},\frac{\alpha_{\theta}}{\alpha_{\theta}+n})$. Here $n_j$ is the number of subjects present in the $j^{th}$ $y$-cluster in the cluster membership in that posterior sample. Similarly $\alpha_{\theta}$ is the updated value from that particular posterior sample. If the value of the first level cluster from the above Multinomial distribution is $k+1$ (that means, a new first level cluster), then we take the values of both the second and third level clusters as 1. In this case, the $p^{th}$ discrete covariate $l^d_p$ is sampled from Bernoulli $(g_p)$, where $g_p\sim$ Beta $(a_0,b_0)$, for $p=1,2,\cdots,p_1$. Also the $p^{th}$ continuous covariate $l^c_p$ is sampled from N$(h_p,\sqrt{f_p})$, where $f_p\sim$ Inv-Chisq $(\nu_0,\tau_0)$, $h_p\sim$ N$(\mu_0,\sqrt{\frac{f_p}{c_0}})$ and $p=1,2,\cdots,p_2$. 
    \vspace{0.2in}\\
     \textit{Substep a.2}: If the value of the first level cluster from the above Multinomial is $j < k+1$ (that means, one of the existing $k$ first level clusters), then we draw the second level cluster from a Multinomial $\{1,2,\cdots,k_j+1\}$ with the probabilities  $(\frac{n_{1\vert j}}{\alpha_{\omega}+n_{j}},\frac{n_{2\vert j}}{\alpha_{\omega}+n_{j}},\cdots,\frac{n_{k_j\vert j}}{\alpha_{\omega}+n_{j}},\frac{\alpha_{\omega}}{\alpha_{\omega}+n_{j}})$. Here $n_{l\lvert j}$ is the number of subjects present in the $l^{th}$ $m$-cluster within the $j^{th}$ $y$-cluster in that iteration, and also $\alpha_{\omega}$ is the update from that particular iteration. As defined in Section \ref{comp_cau}, $k_j$ is the number of non-empty m-clusters within the $j^{th}$ y-cluster in that posterior sample. If the value of the second level cluster drawn from the above Multinomial is $k_j+1$ (that means a new second level cluster within the $j^{th}$ first level cluster), then we take the value of the third level cluster as 1. In this case, we draw $\{l^d_p\}_{p=1}^{p_1}$ and $\{l^c_p\}_{p=1}^{p_2}$ in the aforementioned way. 
     \vspace{0.2in}\\
     \textit{Substep a.3}: If the value of the second level cluster is $l < k_j+1$ (that means, one of the existing $k_j$ second level clusters within the $j^{th}$ first level cluster), then we draw the third level cluster from a Multinomial $\{1,2,\cdots,k_{jl}+1\}$ with probabilities $(\frac{n_{1\vert jl}}{\alpha_{\psi}+n_{l\vert j}},\frac{n_{2\vert jl}}{\alpha_{\psi}+n_{l \vert j}},\cdots,\frac{n_{k_{jl}\vert jl}}{\alpha_{\psi}+n_{l \vert j}},\frac{\alpha_{\psi}}{\alpha_{\psi}+n_{l \vert j}})$. Here $n_{u \lvert jl}$ is the number of subjects present in the $u^{th}$ $x$-cluster of the $l^{th}$ $m$-cluster within the $j^{th}$ $y$-cluster in that particular iteration and $\alpha_{\psi}$ is the update from that iteration. As mentioned in Section \ref{comp_cau}, $k_{jl}$ is the number of non-empty $x$-cluster within the $l^{th}$ $m$-cluster of $j^{th}$ $y$-cluster in that iteration. If the value of the third level cluster is $k_{jl}+1$ (that means, a new third level cluster within the $l^{th}$ second level cluster of $j^{th}$ first level cluster), then we draw $\{l^d_p\}_{p=1}^{p_1}$ and $\{l^c_p\}_{p=1}^{p_2}$ in the similar way as above. 
     \vspace{0.2in}\\
     \textit{Substep a.4}: If the value of the third level cluster is less than $k_{jl}+1$ ( that means, one of the existing $k_{jl}$ third level cluster within the $l^{th}$ second level cluster of $j^{th}$ first level cluster), we then draw $\{l^d_p\}_{p=1}^{p_1}$ and $\{l^c_p\}_{p=1}^{p_2}$, but now the parameters $g_p$, $h_p$ and $f_p$ for the pertinent clusters are taken as the update from that iteration of the Gibbs sampler (as opposed to drawing them from the priors in all the previous cases, where a new third level cluster was drawn). We use the notation $l^{t}$ to denote the vector of covariates drawn in this step, that is, $l^{t}=[\{l_p^d\}_{p=1}^{p_1},\{l_p^c\}_{p=1}^{p_2}]$. This notation will be used in steps b and c).\\
    \item \textit{\textbf{Step b}}: Given the covariates $l^{t}$, drawn in step (a), we now draw the $m$-cluster and the mediators. First we fix a value of $\{a_1,a_2,\cdots,a_Q\}\in\{0,1\}$, which specifies whether the $q^{th}$ mediator is induced under the treatment (that means, $a_q=1$) or not ($a_q=0$). We then separately draw two $m$-clusters from the following Multinomial Distribution; one for the mediators that are induced under the treatment ($a=1$ below), and the other for the mediators that are not induced ($a=0$ below).
    \begin{itemize}
        \item[--] Draw a completely new $m$-cluster with probability $\frac{\alpha_\omega}{\alpha_\omega+n^*} \text{ }k_0 (a,l^{t})$\\
        \item[--]Select existing $j^{th}$ $m$-cluster, but new $x$-cluster within that, with probability $\frac{n_j^*}{\alpha_\omega+n^*}\text{ }\frac{\alpha_\psi}{\alpha_\psi+n_j^*}\text{ }k_0(a,l^{t})$\\
        \item[--]Select existing $j^{th}$ $m$-cluster and existing $l^{th}$ $x$-cluster within that, with probability $\frac{n_j^*}{\alpha_\omega+n^*}\text{ }\frac{n^*_{l\lvert j}}{\alpha_\psi+n_j^*}\text{ }k(a,l^{t})$\\ 
    \end{itemize}
Here $n^*$ is the same as $n$ in the previous step, $n^*_j$ is the number of subjects present in the $j^{th}$ $m$-cluster of that iteration. Similarly, $n^*_{l \lvert j}$ is the number of subjects present in the $l^{th}$ $x$-cluster of $j^{th}$ $m$-cluster in that iteration. $k(\cdot)$ is the probability distribution of the covariates, and $k_0(\cdot)$ is the distribution after integrating the parameters over the prior distribution. As in Step a, $\alpha_{\omega}$ and $\alpha_{\psi}$ are the updates from that particular posterior sample. Once the $m$-clusters are determined under both $a=1$ and $a=0$, we then draw the mediators using the cluster-specific local regression discussed in Section \ref{mod_bnp}. Note that, while drawing the $q^{th}$ mediator, we take the $m$-cluster drawn above with $a=1$ (or, $a=0$), if that mediator is induced (or, not induced) under the treatment, and also the treatment status is taken as $a_q$ in the cluster-specific local regression. Finally, it is worth mentioning that, when the mediators are drawn from the existing $m$-clusters (the last two of the above three cases), the parameters of the cluster-specific local regressions are given by the Gibbs sampler. On the other hand, when the mediators are drawn from a new $m$-cluster (the first one of the above three cases), the local regression parameters for that new cluster is drawn from the prior distribution. We denote the mediators drawn from this step as $m^t$.\\
     \item \textit{\textbf{Step c}}: Given $l^{t}$ and $m^{t}$, we compute $E(Y \vert A=a, L=l^t, M=m^t, \theta^*, \omega^*, \psi^*,s)$ as follows:
\vspace{1\baselineskip}
\begin{align*}
\label{Expeceq} 
&E(Y \vert A=a, L=l^t, M=m^t, \theta^*, \omega^*, \psi^*,s)\\
\vspace{0.2in}\\
=&\frac{w_{k+1}(a,l^t,m^t)E_0(y \vert a,l^t,m^t)+ \sum_{j=1}^k w_{j}(a,l^t,m^t)E(y \vert a,l^t,m^t, \theta^*_j)}{w_{k+1}(a,l^t,m^t)+\sum_{j=1}^k w_{j}(a,l^t,m^t)}
\end{align*}
\vspace{1\baselineskip} 
where $w_{k+1}(a,l^t,m^t)= \frac{\alpha_{\theta}}{\alpha_{\theta}+n} k_0(a,l^t,m^t)$ and 
\begin{align*}
w_{j}(a,l^t,m^t)&= \frac{n_j}{\alpha_{\theta}+n}\{\frac{\alpha_{\omega}}{\alpha_{\omega}+n_j} k_0(m^t \vert a,l^t) k_0(a,l^t)\\
&+ \sum_{l=1}^{k_j}\frac{n_{l\vert j}}{\alpha_{\omega}+n_j}\frac{\alpha_{\psi}}{\alpha_{\psi}+n_{l \vert j}}k(m^t \vert a,l^t;\omega^*_{l \vert j}) k_0(a,l^t)\\
&+ \sum_{l=1}^{k_j}\frac{n_{l\vert j}}{\alpha_{\omega}+n_j} \sum_{u=1}^{k_{jl}}\frac{n_{u\vert jl}}{\alpha_{\psi}+n_{l \vert j}}k(m^t \vert a,l^t;\omega^*_{l \vert j}) k(a,l^t;\psi^*_{u \vert j,l})\}
\end{align*} 
As in \cite{roy2018bayesian}, the terms involving $k_0$ and $E_0$ characterize the distribution and mean respectively, after integrating the parameters over the prior distributions. As defined in Section \ref{comp_cau}, the notation $\theta^*$, $\omega^*$ and $\psi^*$ denote $\{\theta^*_j\}_{j}$, $\{\omega^*_{l\lvert j}\}_{j,l}$ and $\{\psi^*_{u\lvert j,l}\}_{j,l,u}$ respectively for that posterior sample, and $s=\{s_i\}_{i=1}^n$ denotes the cluster memberships of the subjects from that posterior sample.\\
\item \textit{\textbf{Step d}}: In this step, we integrate over the distribution of $(L,M)$ using MC integration. In particular, we repeat the above three steps $T$ times, and for each $t=1,2,\cdots,T$, we compute $E(Y \vert A=a, L= l^t, M=m^t,\theta^*, \omega^*,\psi^*, s)$ and then compute
$$
E(Y(a,M(a_1,a_2,\cdots,a_Q)))\approx \frac{1}{T} \sum_{t=1}^T E(Y \vert A=a, L= l^t, M=m^t,\theta^*, \omega^*,\psi^*, s).
$$
\end{itemize}
\section{Proofs}
\label{Proofs}
\subsection{Proof of Theorem \ref{thm_ident}}
\label{ident_proof}
We first start with identification of NDE. For ease of illustration, we assume that there are three mediators. Conditional on the confounders $L=l$, NDE is identified as\\
\begin{align*}
    &NDE(l)\\
    =& E[Y(1,M(0,0,0))-Y(0,M(0,0,0))\lvert L=l]\\
    =& \int E[Y(1,M(0,0,0))\lvert M(1,1,1)=m_1, M(0,0,0)=m_0, L=l]\\
    &dF_{M(0,0,0),M(1,1,1)\lvert L=l}(m_0,m_1) -E[Y(0,M(0,0,0))\lvert L=l]\\
    =& \int E[Y(1,M(1,1,1))\lvert M(1,1,1)=m_0,L=l]dF_{M(0,0,0)\lvert L=l}(m_0)\\
    &-E[Y(0,M(0,0,0))\lvert L=l], \text{ by Assumption } \ref{A2}.\\
    &\\
    &\text{Now using Assumption \ref{A1}, the above can be written as}\\
    =& \int E[Y\lvert A=1, m_0, L]dF(m_0 \lvert A=0, L)-E[Y\lvert A=0, L]\\
\end{align*}
The terms in the above expression can be identified from the observed data distribution and thus the identifiability follows for NDE.\\
\\
Next, conditional on $L=l$, JNIE is identified as
\begin{align*}
    &JNIE(l)\\
    =& E[Y(1,M(1,1,1))-Y(1,M(0,0,0))\lvert L=l]\\
    =&E[Y(1,M(1,1,1))\lvert L=l]\\ & -\int E[Y(1,M(0,0,0))\lvert M(1,1,1)=m_1, M(0,0,0)=m_0, L=l]\\
    &dF_{M(0,0,0),M(1,1,1)\lvert L=l}(m_0,m_1) \\
    =&E[Y(1,M(1,1,1))\lvert L=l]\\ 
    & -\int E[Y(1,M(1,1,1))\lvert M(1,1,1)=m_0, L=l] dF_{M(0,0,0)\lvert L=l}(m_0)\\
    &\text{by Assumption \ref{A2}}.\\
    &\text{Now using Assumption \ref{A1}, the above can be written as}\\
    &=E[Y\lvert A=1, L]-\int E[Y\lvert A=1, m_0, L]dF(m_0 \lvert A=0, L)\\
\end{align*} 
As before, the terms in the above expression can be identified from the observed data distribution and thus the identifiability follows for JNIE.
\vspace{0.1in}\\
Finally, we prove the identifiability of $\text{NIE}_1$. Identifiability of all the other components of the JNIE will follow similarly. Note that, conditional on $L=l$, $\text{NIE}_1$ can be estimated as
\begin{align*}
    &NIE_1(l)\\
    =& E[Y(1,M(1,1,1))-Y(1,M(0,1,1))\lvert L=l]\\
    =&E[Y(1,M(1,1,1))\lvert L=l]\\ & -\int E[Y(1,M(0,1,1))\lvert M(1,1,1)=m_1, M(0,0,0)=m_0, L=l]\\
    &dF_{M(0,0,0),M(1,1,1)\lvert L=l}(m_0,m_1) \\
    =&E[Y(1,M(1,1,1))\lvert L=l]\\ 
    & -\int E[Y(1,M(1,1,1))\lvert M(1,1,1)=m_{011}, L=l] dF_{M(0,1,1)\lvert L=l}(m_{011})\\
    &\text{by Assumption \ref{A2}}.\\
    &\\
    &\text{Now, by Assumption \ref{A1}, the first term can be written as $E[Y\lvert A=1, L]$ and the integrand}\\ 
    &\text{in the second term can be written as $E(Y \lvert A=1, m_{011}, L)$. Finally, by Assumption \ref{A3}}\\ 
    &\text{and Assumption \ref{A1}, the joint distribution of $M(0,1,1)$ in the second term can be written as}\\
    &\text{$F(M_1 \lvert A=0,L)\times F(M_2,M_3\lvert A=1,L)$. Thus the above equation is equal to the following}\\
    &\\
    &=E[Y\lvert A=1, L]-\int E(Y \lvert A=1, m_{011}, L) d\{F(M_1 \lvert A=0,L)\times F(M_2,M_3\lvert A=1,L)\}\\
\end{align*}
As before, the terms in the above expression can be identified from the observed data distribution and thus the identifiability follows for $\text{NIE}_1$. The remaining mediation effects can be identified analogously.

\subsection{Proof of Theorem \ref{Theo1}}
\label{proof_theo_2}
\begin{proof}
From the aforementioned predictive distribution, it follows that the joint distribution can be expressed as:\\
\\
$Pr(X_1=i_1, M_1=j_1, Y_1=l_1,\cdots, X_n = i_n, M_n = j_n, Y_n = l_n)=\\
\\
\prod_{t=1}^n \frac{\alpha(i_t) + \sum_{h=1}^{(t-1)} \delta_{i_h}(i_t)}{\alpha(\Xset)+t-1} \cdot \frac{\mu(j_t,i_t) + \sum_{h=1}^{(t-1)} \delta_{j_h,i_h}(j_t,i_t)}{\mu(\Mset,i_t)+ \sum_{h=1}^{(t-1)} \delta_{i_h}(i_t)} \cdot \frac{\gamma(l_t,j_t,i_t) + \sum_{h=1}^{(t-1)} \delta_{l_h,j_h,i_h}(l_t,j_t,i_t)}{\gamma(\Yset,j_t,i_t)+ \sum_{h=1}^{(t-1)} \delta_{j_h,i_h}(j_t,i_t)}\cdot$\\
\\
An equivalent representation of the above expression is as follows:  
\begin{dmath}\label{eq1}
Pr(X_1=i_1, M_1=j_1, Y_1=l_1,\cdots, X_n = i_n, M_n = j_n, Y_n = l_n)=p_\alpha \times p_\mu \times p_\gamma\text{ },
\end{dmath}
where\\
\begin{itemize}
\item $p_\alpha=\frac{\Gamma(\alpha(\Xset))}{\prod_{i=1}^k \Gamma(\alpha(i))} \cdot \frac{\prod_{i=1}^k \Gamma(\alpha(i)+n_i)}{\Gamma(\alpha(\Xset)+n)} $
\item $p_\mu=\prod_{i=1}^k \frac{\Gamma(\mu(\Mset,i))}{\prod_{j=1}^r \Gamma(\mu(j,i))} \prod_{i=1}^k \frac{\prod_{j=1}^r \Gamma(\mu(j,i)+n_{ij})}{\Gamma(\mu(\Mset,i)+n_i)} $
\item $p_\gamma=\prod_{i=1}^k \prod_{j=1}^r \frac{\Gamma(\gamma(\Yset,j,i))}{\prod_{l=1}^s \Gamma(\gamma(l,j,i))} \prod_{i=1}^k \prod_{j=1}^r \frac{\prod_{l=1}^s \Gamma(\gamma(l,j,i)+n_{ijl})}{\Gamma(\gamma(\Yset,j,i)+n_{ij})}\cdot$
\end{itemize}
Looking at the above expression, it is evident that the joint distribution depends on the number of unique triplets seen, not on the order in which they are observed. Hence the triplet $\{X_n, M_n, Y_n\}_{n \in \Natural}$ form an exchangeable sequence. By de Finetti's Representation Theorem, there exists a probability measure $\tilde{Q}$ on the simplex $S_{k,r,s}=\{p_{1,1,1}, \cdots,p_{k,r,s}: p_{i,j,l} \geq 0$ and $\sum_{i=1}^k \sum_{j=1}^r \sum_{l=1}^s p_{i,j,l}=1 \}$ such that \\
\\
$Pr(X_1=i_1, M_1=j_1, Y_1=l_1,\cdots, X_n = i_n, M_n = j_n, Y_n = l_n)\\
\\
=\int_{[0,1]^{krs}}\prod_{i=1}^k \prod_{j=1}^r \prod_{l=1}^s p_{i,j,l}^{n_{ijl}}\tilde{Q}(dp_{1,1,1},\cdots, dp_{k,r,s}).$\\
\\
Define the simplices $S_{k}=\{p_{1}, \cdots,p_{k}: p_{i} \geq 0$ and $\sum_{i=1}^k p_{i}=1 \}$, $S_{r}^{(i)}=\{p_{1 \vert i}, \cdots,p_{r \vert i}: p_{j \vert i} \geq 0$ and $\sum_{j=1}^r p_{j \vert i}=1 \}$ for $i=1,2,\cdots,k$ and $S_{s}^{(i,j)}=\{p_{1 \vert (i,j)}, \cdots,p_{s \vert (i,j)}: p_{l \vert (i,j)} \geq 0$ and $\sum_{l=1}^s p_{l \vert (i,j)}=1 \}$ for $i=1,2,\cdots,k$ and $j=1,2,\cdots,r$. Let $Q$ be the probability measure on the product of the simplexes $S_k \times \prod_{i=1}^k S_r^{(i)} \times \prod_{i=1}^k \prod_{j=1}^r S_s^{(i,j)}$, obtained from $\tilde{Q}$ via a reparameterization in terms of $(p_1,\cdots, p_k,p_{1 \vert 1},\cdots, p_{r \vert k},p_{1\vert (1,1)},\cdots,p_{s\vert (k,r)})$  . Then, 
\begin{dmath}\label{eq2}
Pr(X_1=i_1, M_1=j_1, Y_1=l_1,\cdots, X_n = i_n, M_n = j_n, Y_n = l_n)=\\
 \int_{[0,1]^{k} \times [0,1]^{kr} \times [0,1]^{krs}}\prod_{i=1}^k 
p_i^{n_i} \prod_{j=1}^r p_{j \vert i}^{n_{ij}} \prod_{l=1}^s p_{l\vert(i,j)}^{n_{ijl}}Q(dp_{1},\cdots, dp_{s\vert(k,r)}).
\end{dmath}
Now, since the Dirichlet distribution is determined by its moments, combining equation \ref{eq1} and equation \ref{eq2}, one can conclude that,
\begin{center}
\begin{align*}
p_1,\cdots,p_k &\sim Dir(\alpha(1),\cdots,\alpha(k))\\
p_{1\vert i},\cdots,p_{r\vert i} &\sim Dir(\mu(1,i),\cdots,\mu(r,i)), i=1,2,\cdots,k\\
p_{1\vert(i,j)},\cdots,p_{s\vert(i,j)} &\sim Dir(\gamma(1,j,i),\cdots,\gamma(s,j,i)), i=1,2,\cdots,k; j=1,2,\cdots,r  
\end{align*}
\end{center}
where $(p_1,\cdots,p_k)$, $(p_{1\vert i},\cdots,p_{r\vert i})$ and $(p_{1\vert(i,j)},\cdots,p_{s\vert(i,j)})$ are independent.
\paragraph{}
The second part of the theorem follows from de Finetti's result on asymptotic behavior of the predictive distributions for exchangeable sequences. See \cite{cifarelli1996finetti} for more details. 
\end{proof}
\subsection{Proof of Theorem \ref{Theo_post_cons}}
\begin{proof}
First, we start with $E[Pr(A\times B\times C) \lvert X_1 = x_1, M_1=m_1, Y_1=y_1,\cdots,X_n=x_n, M_n=m_n, Y_n =y_n]$. We divide the range of the integration over two regions: $(A\times B)\setminus \{(x_1,m_1), (x_2,m_2), \cdots,(x_n,m_n)\}$ and $(A\times B)\cap \{(x_1,m_1), (x_2,m_2), \cdots,(x_n,m_n)\}$ and arrive at the two terms of the above expectation. The first term from the first integration will have the following form, which is induced by the base distribution : $\frac{\alpha(\Xset)}{\alpha(\Xset)+n}P_0((A\times B)\setminus \{(x_1,m_1), (x_2,m_2), \cdots,(x_n,m_n)\} \times C)$. The second term of the expectation has the following form:\\
\\
\begin{align*}
&\int_{(A\times B)\cap \{(x_1,m_1), (x_2,m_2), \cdots,(x_n,m_n)\}} \frac{\gamma(C,m,x)+\sum_{k=1}^{n_{mx}}\delta_{y (m x,k)}(C)}{\gamma(\Yset,m,x)+n_{mx}}d\left(\frac{\mu(m,x)+\sum_{j=1}^{n_x}\delta_{m (x,j)}(m)}{\mu(\Mset,x)+n_x}\frac{\alpha(x)+\sum_{i=1}^n\delta_{x_{i}}(x)}{\alpha(\Xset)+n}\right)\\
&\\
&=\sum_{(x,m) \in (A \times B) \cap \{(x_1,m_1),\cdots,(x_n,m_n)\}}\frac{\gamma(C,m,x)+\sum_{k=1}^{n_{mx}}\delta_{y (m x,k)}(C)}{\gamma(\Yset,m,x)+n_{mx}}\cdot\frac{\mu(m,x)+n_{mx}}{\mu(\Mset,x)+n_x}\cdot\frac{\alpha(x)+n_x}{\alpha(\Xset)+n}.\\
\end{align*}\\
Hence looking at the two terms of the expectation,\\
\begin{align*}
&\\
&E[Pr(A\times B\times C) \lvert X_1 = x_1, M_1=m_1, Y_1=y_1,\cdots,X_n=x_n, M_n=m_n, Y_n =y_n]\\
&\\
&\sim \frac{1}{n}\sum_{(x,m) \in (A \times B) \cap \{(x_1,m_1),\cdots,(x_n,m_n)\}} \sum_{k=1}^{n_{mx}}\delta_{y (m x,k)}(C)\\
&\\
&=\frac{1}{n} \sum_{i=1}^{n} \delta_{x_i,m_i,y_i}(A,B,C)
&\\
&\\
&\xrightarrow{}\pi(A \times B\times C) a.s.\pi^{\infty}.\\
\end{align*}\\
Using a similar calculation as in the proof of Lemma 1 in \cite{wade2011enriched}, we obtain the following:
\begin{align*}
&E[P(A \times B\times C)^2] =\\
&\\
&\frac{1}{\alpha(\Xset)+1} \int_{A\times B}\frac{1}{\mu(\Mset,x)+1} \cdot \frac{P_{0Y\lvert MX}(C\lvert m,x)(1+\gamma(C,m,x)P_{0Y\lvert MX}(C \lvert m,x))}{\gamma(C,m,x)+1} d(P_{0M\lvert X}(m \lvert x) \cdot P_{0X}(x))\\
&\\
&+\frac{\alpha(\Xset)}{\alpha(\Xset)+1} \int_{A\times B} \int_{\{x\}\times\{m\lvert x\}}\frac{\mu(\Mset,x)}{\mu(\Mset,x)+1} \cdot \frac{P_{0Y\lvert MX}(C\lvert m,x)(1+\gamma(C,m,x)P_{0Y\lvert MX}(C \lvert m,x))}{\gamma(C,m,x)+1}\\
&\\
&d(P_{0M\lvert X}(m^ \prime \lvert x^ \prime) \cdot P_{0X}(x ^\prime)) d(P_{0M\lvert X}(m \lvert x) \cdot P_{0X}(x))\\
&\\
&+ \frac{\alpha(\Xset)}{\alpha(\Xset)+1} \int_{A\times B} \int_{\{x\}\times(B \setminus \{m\lvert x\})} \frac{\mu(\Mset,x)}{\mu(\Mset,x)+1} \cdot P_{0Y\lvert MX}(C\lvert m^\prime,x^\prime) \cdot P_{0Y\lvert MX}(C\lvert m,x)\\
&\\
&d(P_{0M\lvert X}(m^ \prime \lvert x^ \prime) \cdot P_{0X}(x ^\prime)) d(P_{0M\lvert X}(m \lvert x) \cdot P_{0X}(x))\\
&\\
&+ \frac{\alpha(\Xset)}{\alpha(\Xset)+1} \int_{A\times B} \int_{(A \setminus \{x\})\times(B \setminus \{m\lvert x\})} \frac{\mu(\Mset,x)}{\mu(\Mset,x)+1} \cdot P_{0Y\lvert MX}(C\lvert m^\prime,x^\prime) \cdot P_{0Y\lvert MX}(C\lvert m,x)\\
&\\
&d(P_{0M\lvert X}(m^ \prime \lvert x^ \prime) \cdot P_{0X}(x ^\prime)) d(P_{0M\lvert X}(m \lvert x) \cdot P_{0X}(x))\\
&\\
&+ \frac{\alpha(\Xset)}{\alpha(\Xset)+1} \int_{A\times B} \int_{(A \setminus \{x\})\times \{m\lvert x\}} \frac{\mu(\Mset,x)}{\mu(\Mset,x)+1} \cdot P_{0Y\lvert MX}(C\lvert m^\prime,x^\prime) \cdot P_{0Y\lvert MX}(C\lvert m,x)\\
&\\
&d(P_{0M\lvert X}(m^ \prime \lvert x^ \prime) \cdot P_{0X}(x ^\prime)) d(P_{0M\lvert X}(m \lvert x) \cdot P_{0X}(x)).\\
\end{align*}\\
With some algebra, it is easy to see that,\\
\\
\begin{align*}
&Var(P(A \times B\times C))\\
&\\
&=\frac{1}{\alpha(\Xset)+1} \int_{A\times B}\frac{1}{\mu(\Mset,x)+1} \cdot \frac{P_{0Y\lvert MX}(C\lvert m,x)(1+\gamma(C,m,x)P_{0Y\lvert MX}(C \lvert m,x))}{\gamma(C,m,x)+1} d(P_{0M\lvert X}(m \lvert x) \cdot P_{0X}(x))\\
&\\
&+\frac{\alpha(\Xset)}{\alpha(\Xset)+1} \int_{A\times B} \int_{\{x\}\times\{m\lvert x\}}\frac{\mu(\Mset,x)}{\mu(\Mset,x)+1} \cdot \frac{P_{0Y\lvert MX}(C\lvert m,x)(1+\gamma(C,m,x)P_{0Y\lvert MX}(C \lvert m,x))}{\gamma(C,m,x)+1}\\
&\\
&d(P_{0M\lvert X}(m^ \prime \lvert x^ \prime) \cdot P_{0X}(x ^\prime)) d(P_{0M\lvert X}(m \lvert x) \cdot P_{0X}(x))\\
&\\
&-\frac{1}{\alpha(\Xset)+1} \int_{A\times B} \int_{\{x\}\times\{m\lvert x\}}\frac{1}{\mu(\Mset,x)+1} \cdot P^2_{0Y \lvert MX}(C \lvert m,x) d(P_{0M\lvert X}(m^ \prime \lvert x^ \prime) \cdot P_{0X}(x ^\prime)) d(P_{0M\lvert X}(m \lvert x) \cdot P_{0X}(x))\\
&\\
&- \frac{1}{\alpha(\Xset)+1} \int_{A\times B} \int_{\{x\}\times(B \setminus \{m\lvert x\})} \frac{1}{\mu(\Mset,x)+1} \cdot P_{0Y\lvert MX}(C\lvert m^\prime,x^\prime) \cdot P_{0Y\lvert MX}(C\lvert m,x)\\
&\\
&d(P_{0M\lvert X}(m^ \prime \lvert x^ \prime) \cdot P_{0X}(x ^\prime)) d(P_{0M\lvert X}(m \lvert x) \cdot P_{0X}(x))\\
&\\
&- \frac{1}{\alpha(\Xset)+1} \int_{A\times B} \int_{(A \setminus \{x\})\times(B \setminus \{m\lvert x\})} \frac{1}{\mu(\Mset,x)+1} \cdot P_{0Y\lvert MX}(C\lvert m^\prime,x^\prime) \cdot P_{0Y\lvert MX}(C\lvert m,x)\\
&\\
&d(P_{0M\lvert X}(m^ \prime \lvert x^ \prime) \cdot P_{0X}(x ^\prime)) d(P_{0M\lvert X}(m \lvert x) \cdot P_{0X}(x))\\
&\\
&- \frac{1}{\alpha(\Xset)+1} \int_{A\times B} \int_{(A \setminus \{x\})\times \{m\lvert x\}} \frac{1}{\mu(\Mset,x)+1} \cdot P_{0Y\lvert MX}(C\lvert m^\prime,x^\prime) \cdot P_{0Y\lvert MX}(C\lvert m,x)\\
&\\
&d(P_{0M\lvert X}(m^ \prime \lvert x^ \prime) \cdot P_{0X}(x ^\prime)) d(P_{0M\lvert X}(m \lvert x) \cdot P_{0X}(x)).\\
\end{align*}\\
Now, using the fact that $\frac{\alpha_n(A)}{\alpha_n(\Xset)} \sim \frac{1}{n} \sum_{i=1}^n \delta_{x_i}(A)$, $\frac{\mu_n(B,x)}{\mu_n(\Mset,x)} \sim \frac{1}{n_x} \sum_{j=1}^{n_x}\delta_{m(x,j)}(B)$, $\frac{\gamma_n(C,m,x)}{\gamma_n(\Yset,m,x)}\sim \frac{1}{n_{mx}} \sum_{k=1}^{n_{mx}}\delta_{y(mx,k)}(C)$ and following similar steps as in the proof of Theorem 6 of \cite{wade2011enriched}, it can be shown that the posterior variance of $P(A\times B\times C)$ goes to $0$. Hence the result follows.
\end{proof}
\section{Additional Tables from Sections \ref{simul} and \ref{application}}
\label{add_tables}
\begin{table}
\centering
\begin{tabular}{|c | c| c| c| c|}
\hline
\textit{Mediator No.} & \textit{True INIE} & \textit{Estimate} & \textit{CI Width} & \textit{Coverage}\\
\hline
1&0&0.0002 &0.25 &0.99 \\
2&0&0.005 &0.24 &0.99 \\
3&0&-0.001 & 0.26&0.98 \\
4&0&0.003 &0.24 &0.99 \\
5&0&-0.003& 0.26&0.97 \\
6&0&0.0008& 0.25&0.99 \\
7&0& -0.004&0.24 &0.98 \\
8&0&-0.001&0.23&0.97 \\
9&0&0.018 &0.25 &0.99 \\
10&0.70&0.71&0.32 &0.97 \\
\hline
\end{tabular}
\caption{Scenario 1 INIE results for $n=1000$} \label{scn_1_INIE}
\end{table}
\begin{table*}
\centering
\begin{tabular}{c c c c c}
    \multicolumn{2}{c}{}& \multicolumn{1}{c}{\textit{Estimate}} & \multicolumn{1}{c}{\textit{CI Width}} & \multicolumn{1}{c}{\textit{Coverage}}\\  
    \hline
      \multirow{2}{*}{\textit{True NDE}=$1.51$} & \small{BNP ($n=1000$)} &1.55&1.78&0.96\\
      &\small{BNP ($n=2000$)} &1.53&1.55&0.97\\
      &\small{LSEM ($n=1000$)} &0.93&0.81&0.21\\
      &\small{LSEM ($n=2000$)} &0.94&0.57&0.23\\
      \hline
      \multirow{2}{*}{\textit{True JNIE}=$0.41$} & \small{BNP ($n=1000$)} &0.47&1.71&0.99\\
      &\small{BNP ($n=2000$)} &0.42&1.51&0.99\\
      &\small{LSEM ($n=1000$)} &0.95&0.46&0.46\\
      &\small{LSEM ($n=2000$)} &0.93&0.32&0.46\\
       \hline
       \multirow{2}{*}{\textit{True TE}=$1.92$} & \small{BNP ($n=1000$)} &2.01&0.90&0.99\\
       &\small{BNP ($n=2000$)} &1.96&0.74&0.99\\
      &\small{LSEM ($n=1000$)} &1.88&0.89&0.88\\
      &\small{LSEM ($n=2000$)} &1.89&0.63&0.89\\
       \hline
    \end{tabular}
    \caption{Scenario 2 results for NDE, JNIE, TE}
    \vspace{0.8cm}
    \label{Sim_scn2}
\end{table*}
\begin{table}
\centering
\begin{tabular}{|c | c| c| c| c|}
\hline
\textit{Mediator No.} & \textit{True INIE} & \textit{Estimate} & \textit{CI Width} & \textit{Coverage}\\
\hline
1&0 & 0.028&0.94 &0.96 \\
2&0&0.012 &0.96 &0.96 \\
3&0& 0.009& 0.95&0.96 \\
4&0&0.03 &0.94 &0.97 \\
5&0& 0.018& 0.97&0.95 \\
6&0& 0.005& 0.95&0.97 \\
7&0& 0.025& 0.96&0.97 \\
8&0& 0.038&0.95 &0.97 \\
9&0&0.027 &0.95 &0.96 \\
10&0.41&0.35&0.95&0.96 \\
\hline
\end{tabular}
\caption{Scenario 2 INIE results for $n=1000$} \label{scn_2_INIE}
\end{table}
\begin{table}
\centering
\begin{tabular}{|c | c| c| c| c|}
\hline
\textit{Mediator No.} & \textit{True INIE} & \textit{Estimate} & \textit{CI Width} & \textit{Coverage}\\
\hline
1&0 & 0.012&0.83&0.97 \\
2&0&0.008 &0.85&0.97 \\
3&0& 0.001& 0.84&0.98 \\
4&0&0.02 &0.82&0.98 \\
5&0& 0.0078& 0.86&0.97 \\
6&0& 0.005& 0.83&0.98 \\
7&0& 0.013& 0.83&0.97 \\
8&0& 0.017&0.85 &0.98 \\
9&0&0.014 &0.84 &0.97 \\
10&0.41&0.39&0.83&0.98 \\
\hline
\end{tabular}
\caption{Scenario 2 INIE results for $n=2000$} \label{scn_22_INIE}
\end{table}
\vspace{0.6cm}
\begin{table}
\centering
\begin{tabular}{|c | c| c| c| c|}
\hline
\textit{Mediator No.} & \textit{True INIE} & \textit{Estimate} & \textit{CI Width} & \textit{Coverage}\\
\hline
1&0 &0.02 &0.97 &0.99 \\
2&0& 0.02&0.99 &0.99 \\
3&0&0.02 &0.98& 0.98\\
4&0&0.03 &0.96 &0.99 \\
5&0&0.011 &0.97 &0.98 \\
6&0&0.01& 0.99&0.99 \\
7&0& 0.012& 0.98&0.98 \\
8& -2.47&-2.33 &0.97 &0.97 \\
9&2.65&2.51&0.98&0.98 \\
10&5.68&5.53 &0.97 &0.98 \\
\hline
\end{tabular}
\caption{Scenario 3 INIE results for $n=2000$} \label{scn_33_INIE}
\end{table}
\begin{table*}
\centering
\begin{tabular}{c c c c c}
    \multicolumn{2}{c}{}& \multicolumn{1}{c}{\textit{Estimate}} & \multicolumn{1}{c}{\textit{CI Width}} & \multicolumn{1}{c}{\textit{Coverage}}\\  
    \hline
      \multirow{2}{*}{\textit{True NDE}=$0.84$} & \small{BNP ($n=1000$)} &0.71&1.33&0.92\\
      &\small{BNP ($n=2000$)} &0.75&0.90&0.94\\
      &\small{LSEM ($n=1000$)} &1.49&0.75&0.32\\
      &\small{LSEM ($n=2000$)} &1.48&0.58&0.32\\
      \hline
      \multirow{2}{*}{\textit{True JNIE}=$1.04$} & \small{BNP ($n=1000$)} &1.35&1.28&0.97\\
      &\small{BNP ($n=2000$)} &1.32&0.82&0.98\\
      &\small{LSEM ($n=1000$)} &0.51&0.43&0.58\\
      &\small{LSEM ($n=2000$)} &0.50&0.35&0.59\\
       \hline
       \multirow{2}{*}{\textit{True TE}=$1.88$} & \small{BNP ($n=1000$)} &2.03&1.06&0.97\\
       &\small{BNP ($n=2000$)} &1.92&0.91&0.98\\
      &\small{LSEM ($n=1000$)} &2.01&0.81&0.75\\
      &\small{LSEM ($n=2000$)} &1.99&0.62&0.75\\
       \hline
    \end{tabular}
    \caption{Scenario 4 results for NDE, JNIE, TE}
    \vspace{0.8cm}
    \label{Sim_scn4}
\end{table*}

\begin{table}
\centering
\begin{tabular}{|c | c| c| c| c|}
\hline
\textit{Mediator No.} & \textit{True INIE} & \textit{Estimate} & \textit{CI Width} & \textit{Coverage}\\
\hline
1&0&0.004 &0.45 &0.97 \\
2&0&0.01&0.44 &0.98 \\
3&0&0.002&0.48&0.99\\
4&0&-0.021&0.46&0.98 \\
5&0&0.008&0.44&0.98 \\
6&0&0.018 &0.49 &0.99 \\
7&0&0.016&0.47&0.99 \\
8&0&0.015&0.49&0.97\\
9&0&0.022&0.48&0.97\\
10&1.09&1.12&0.45&0.98\\
\hline
\end{tabular}
\caption{Scenario 4 INIE results for $n=1000$} \label{scn_4_INIE}
\end{table}
\begin{table*}
\centering
\begin{tabular}{c c c c c}
    \multicolumn{2}{c}{}& \multicolumn{1}{c}{\textit{Estimate}} & \multicolumn{1}{c}{\textit{CI Width}} & \multicolumn{1}{c}{\textit{Coverage}}\\  
    \hline
      \multirow{2}{*}{\textit{True NDE}=$1.49$} & \small{BNP ($n=1000$)} &1.51&1.74&0.98\\
      &\small{BNP ($n=2000$)} &1.50&1.44&0.99\\
      &\small{LSEM ($n=1000$)} &0.93&0.81&0.28\\
      &\small{LSEM ($n=2000$)} &0.94&0.56&0.28\\
      \hline
      \multirow{2}{*}{\textit{True JNIE}=$0.46$} & \small{BNP ($n=1000$)} &0.48&0.44&1.0\\
      &\small{BNP ($n=2000$)} &0.44&0.43&1.0\\
      &\small{LSEM ($n=1000$)} &0.94&0.45&0.46\\
      &\small{LSEM ($n=2000$)} &0.93&0.32&0.47\\
       \hline
       \multirow{2}{*}{\textit{True TE}=$1.95$} & \small{BNP ($n=1000$)} &1.98&0.88&1.0\\
       &\small{BNP ($n=2000$)} &1.96&0.68&1.0\\
      &\small{LSEM ($n=1000$)} &1.87&0.89&0.89 \\
      &\small{LSEM ($n=2000$)} &1.89&0.64&0.89\\
       \hline
    \end{tabular}
    \caption{Scenario 5 results for NDE, JNIE, TE}
    \vspace{0.5cm}
    \label{Sim_scn5}
\end{table*}
\begin{table}
\centering
\begin{tabular}{|c | c| c| c| c|}
\hline
\textit{Mediator No.} & \textit{True INIE} & \textit{Estimate} & \textit{CI Width} & \textit{Coverage}\\
\hline
1&0& 0.02& 0.88&0.98 \\
2&0& 0.03& 0.85& 0.97\\
3&0&0.02&0.87& 0.98\\
4&0&0.02 &0.86 &0.98 \\
5&0& 0.04& 0.84&0.97 \\
6&0&0.009 &0.86 &0.97 \\
7&0&0.03 &0.84 &0.98 \\
8&0&0.02 &0.87 &0.97 \\
9&0& 0.012&0.85 &0.97 \\
10&0.41&0.38&0.87&0.98\\
\hline
\end{tabular}
\caption{Scenario 5 INIE results for $n=1000$} \label{scn_5_INIE}
\end{table}
\begin{table}
\centering
\begin{tabular}{|c | c| c| c| c|}
\hline
\textit{Mediator No.} & \textit{True INIE} & \textit{Estimate} & \textit{CI Width} & \textit{Coverage}\\
\hline
1&0&-0.005&0.33&0.97\\
2&0&-0.004&0.33&0.96\\
3&0&-0.005&0.34&0.97\\
4&0&-0.006&0.33&0.98\\
5&0&-0.009&0.32&0.97\\
6&0&-0.006&0.33&0.97\\
7&0&-0.008&0.31&0.96\\
8&0&-0.005&0.34&0.98\\
9&0&-0.004&0.33&0.97\\
10&-0.02&-0.01&0.32&0.97\\
\hline
\end{tabular}
\caption{Scenario 6 INIE results for $n=100$} \label{scn_6_INIE}
\end{table}
\begin{table}
  \centering
  \begin{tabular}{c|c|c|c}
    \toprule
    Mediator & INIE Estimate & Lower CI& Upper CI\\
    \midrule
    Vancomycin IV&0.03&-0.21&0.24\\
Metronidazole&-0.006&-0.21&0.22\\
Cefazolin&-0.014&-0.23&0.19\\
Daptomycin&0.01&-0.19&0.21\\
Linezolid & 0.03 & -0.24 & 0.2\\
Meropenem & -0.01 & -0.19  & 0.21\\
Pip Tazo &0.001 & -0.23 & 0.22\\
Cefepime & -0.004 &-0.22 & 0.22\\
Levofloxacin &-0.003 & -0.2 & 0.2\\
Colistimethate &-0.008 & -0.2 & 0.2\\
Clindamycin & 0.013 & -0.2 & 0.24\\
Ceftriaxone & 0.001 & -0.21 &0.21\\
Azithromycin & 0.001 & -0.21 & 0.21\\
Ampsul & 0.004 & -0.2 & 0.2\\
Amikacin & 0.004 & -0.22 & 0.24\\
\bottomrule
\end{tabular}
\caption{INIE for the mediators: Posterior means and $95\%$ Credible Intervals}
\label{data_ana_tab_2}
\end{table}
\clearpage
\bibliographystyle{chicago}
\bibliography{EDP}
\end{document}